\documentclass[12pt,preprint]{aastex}

\def\apgt{\ {\raise-.5ex\hbox{$\buildrel>\over\sim$}}\ }
\def\aplt{\ {\raise-.5ex\hbox{$\buildrel<\over\sim$}}\ }
\def\Msun{M_\odot}

\shorttitle{The Parameter Space of Nova Outbursts}
\shortauthors{Yaron, Prialnik, Shara, Kovetz}

\begin{document}

\title{An Extended Grid of Nova Models: II. The Parameter Space of Nova Outbursts}

\author{
O.~Yaron,\altaffilmark{1}
D.~Prialnik,\altaffilmark{1}
M.~M.~Shara\altaffilmark{2}
and
A.~Kovetz\altaffilmark{1,3}
}

\altaffiltext{1}{Department of Geophysics and Planetary Sciences, 
Sackler Faculty of Exact Sciences, Tel Aviv University, Ramat Aviv 
69978, Israel}
\altaffiltext{2}{Department of Astrophysics, American Museum of Natural
  History, Central Park West and 79th street, New York, NY 10024-5192}
\altaffiltext{3}{School of Physics and Astronomy, Sackler Faculty of
  Exact Sciences, Tel Aviv University, Ramat Aviv 69978, Israel}

\begin{abstract}

This paper is a sequel to an earlier paper devoted to multiple, multicycle nova evolution models 
(\citet{PK95}, first paper of the series), which showed that the different 
characteristics of nova outbursts can be reproduced by varying the values of three basic and
independent parameters: the white dwarf mass-$M_{WD}$, the temperature 
of its isothermal core-$T_{WD}$ and the mass transfer rate-$\dot{M}$.
Here we show that the parameter space is
constrained by several analytical considerations and find its limiting 
surfaces. Consequently, we extend the grid of multicycle nova evolution models 
presented in Paper I to its limits, adding multicycle nova outburst calculations for a 
considerable number of new parameter combinations.
In particular, the extended parameter space that produces nova eruptions includes low mass transfer 
rates down to $5\times10^{-13}~M_\odot$~yr$^{-1}$, and more models for low $T_{WD}$. Resulting characteristics of these
runs are added to the former parameter combination results, to
provide a full grid spanning the entire parameter space for
Carbon-Oxygen white dwarfs.
The full grid covers the entire range of observed nova characteristics, even
those of peculiar objects, which have not been numerically reproduced
until now. Most remarkably, runs for very low $\dot{M}$ lead to very high values
for some characteristics, such as outburst amplitude $A\gtrsim20$,
high super-Eddington luminosities at maximum, heavy element
abundance of the ejecta $Z_{ej}\approx0.63$ and high ejected masses
$m_{ej}\approx 7\times10^{-4}~M_\odot$.

\end{abstract}


\section{INTRODUCTION}
\label{introduction}

The realization that Classical Novae (CN) are due to thermonuclear
runaways (TNR) on the surfaces of white dwarf (WD) stars in close,
mass-transferring binary systems \citep{Sta72} is now three decades 
behind us.
Direct observations of novae (summarized in e.g. \citet{Pay57})
had already demonstrated  that large ranges in ejecta velocity
and metallicity occur in novae, and later interpretations of the data
suggested that ejecta masses varied strongly from object to object. The 
first
simulations by nova theorists, trying to make sense of these
observations, involved much groping into various corners of
parameter space. Early triumphs included an explanation of the UV emission and high 
bolometric luminosities of novae after eruption \citep{Sta76};  fast and moderately
slow nova models \citep{Pri78}; natural enrichment of ejecta  in heavy
elements via diffusion \citep{Pri85}
and via mixing from the underlying white dwarf \citep{Mac83,Kut87};
the realization that nuclear burning with EUV emission
could occur without mass ejection (perhaps leading to SNIa or super soft
X-ray sources) \citep{Sha77} and novae recurring on timescales as short 
as a few
years \citep{Sta87}.

The first decade of nova simulations crystallized the understanding that
a space of three independent parameters controls the behavior of a 
classical
nova eruption: the white dwarf mass,
the white dwarf temperature (or luminosity) and the mass accretion rate
from the companion \citep{Sha80,Pri95}.
Increased computer power, and codes including better physics
lead to an ever-increasingly sophisticated series of eruption
simulations. Once multi-cycle simulations became possible, the
arbitrariness of initial conditions could be eliminated, and
systematic surveys throughout the three-dimensional space became
a real possibility.
These culminated in an extended grid of multicycle nova evolution models
(\citet{PK95}, hereafter Paper I) which have been extensively used by 
researchers.

While Prialnik \& Kovetz covered virtually all of the $(M_{WD},T_{WD},\dot{M})$ 
space then thought likely to be reached by novae, more 
recent work \citep{HurSha03}
has shown that cataclysmic variables are likely to evolve considerably 
beyond the boundaries of their calculations.
Remarkably, we find (in the present study) that some of these systems (parameter combinations)
still produce nova eruptions. Amongst these are models with extreme
values of $M_{WD}$, $T_{WD}$
and $\dot{M}$ which can give rise to "extreme novae" --  objects with
eruption luminosities, metallicities, and/or ejecta masses and
velocities significantly  larger
\emph{or} significantly smaller than any yet predicted by simulations.

The existence of very unusual eruptive systems like the "Red Variable"
in M31 \citep{Mou00} or V838 Mon \citep{Bon03}
are already sufficient justification for extending nova simulation grids
to cover all conceivable cases. Each of these
unusual variables has properties (particularly luminosity) reminiscent
of novae,
and we will examine possible connections in a future paper. But of equal
interest is the physical insight that these new simulations can
provide, as important input in trying to determine the very long term
evolution of cataclysmic binaries,
and relationships between the various subclasses of CVs. Most surprising
of all is the non-linear behavior of
some of the most extreme new nova models. Simple extrapolations from
previous simulations give answers that may be
wrong by large factors. {\it This is because dominance of competing
timescales and physical effects changes with movement
through the three-dimensional nova parameter space.}

It is perhaps not too surprising that $\dot{M}$ lower than 
$10^{-11}~M_\odot$~yr$^{-1}$
(the lowest values considered in Paper I) can occur, especially as a mass-accreting
WD whittles down its donor companion to  the brown dwarf mass
range.  Equally understandable
is that WD core temperatures can drop to or below $10^7$~K, for
WDs accreting at very low
rates for many Gyr, whereas Paper I considered only hot WDs accreting at the
lowest rate. Both these effects - low mass accretion rate and
cold WDs - are extensively studied in the present work.
What is less intuitively obvious is that CO WDs with masses well under 
the $0.65~M_\odot$ (the lower limit of  the Paper I study) are also
expected. Mass stripping via common envelope evolution can produce a
naked helium star, which evolves to the helium giant branch and then
becomes a low-mass CO WD, as illustrated by the example of a long-term
binary evolution track summarized in Table~\ref{tbl:bin-evol} 
(Hurley, private communication). The existence of low-mass ($\lesssim0.50~M_\odot$) WDs 
resulting from close binary evolution was also discussed by \citet{Mar95} and
 \citet{Alt04} (and references therein).
The behavior of such low-mass CO WDs accreting hydrogen-rich material is also examined, 
for the first time, in the present work.

The methods of numerical computation, the grid of results, and the range
of nova parameter space covered are discussed in $\S~2$.
In $\S~3$ we discuss the existence of a confined region of the 
three-dimensional parameter space within which conditions for nova 
outbursts are satisfied. A brief summary and conclusions are given in $\S~4$.

\section{NUMERICAL COMPUTATIONS} \label{numerical}

The hydrodynamic Lagrangian stellar evolution code by which the
present study was performed is the
code presented in Paper I. It includes OPAL opacities, an
extended nuclear reactions network comprising of 40 heavy element
isotopes and a mass-loss algorithm that applies a steady, optically
thick supersonic wind solution (following the phase of rapid
expansion). In addition, diffusion is computed for all elements,
accretional heating is taken into account and convective fluxes are
calculated according to the mixing length theory. Initial models were
prepared for four WD-mass values and three temperatures by cooling
WD models from higher temperatures. Each nova model was followed through
several consecutive outburst cycles in order to eliminate the effect of
the initial configuration. One typical cycle was then chosen as 
representative of each parameter combination.


The original grid presented in Paper I consisted of $64$ parameter
combinations: four $M_{WD}$ values---$0.65,1.00,1.25\ \&\ 1.40~M_\odot$,
three $T_{WD}$ values---$10,30\ \&\ 50\ \times10^6$~K
and five $\dot{M}$ values---$10^{-6}$ through $10^{-10}~M_\odot$~yr$^{-1}$,
and four calculations for the highest temperature value with $\dot{M}$ of
$10^{-11}~M_\odot$~yr$^{-1}$. In the present work we have extended the 
grid to lower accretion rates, down to
$\dot{M}=5\times10^{-13}~M_\odot$~yr$^{-1}$ for all values of WD mass and
temperature and we have added calculations for $M_{WD}=0.4~M_\odot$,
amounting to about 30 new parameter combinations in all.

Tables~\ref{tbl:envelope} \& \ref{tbl:outburst} list the results of
our complete grid of parameter combinations.
The tables are presented in a manner similar to that of Paper I:
Table~\ref{tbl:envelope} lists properties that are related to the
accretion phase and the onset of the outburst, whereas
Table~\ref{tbl:outburst} presents characteristics of the outburst
itself. The properties displayed in Table~\ref{tbl:envelope} are: the
accreted and ejected masses $m_{acc},\ m_{ej}$; the helium mass
fraction of the envelope and the ejecta
$Y_{env},\ Y_{ej}$; the heavy element mass fractions $Z_{env},\
Z_{ej}$; and the maximum temperature attained in the burning shell
$T_{max}$. The characteristics displayed in
Table~\ref{tbl:outburst} are: the maximal expansion velocity
$v_{max}$, its average over the whole mass-loss phase $v_{av}$,
the maximal luminosity attained during
the outburst $L_{4,max}$, the amplitude
of the outburst in magnitudes and three typical timescales---the time of
decline of bolometric luminosity by 3 mag. $t_{3,bol}$, the duration
of the mass-loss phase $t_{m-l}$, and the recurrence period of the outbursts
$P_{rec}\equiv m_{acc}/\dot{M}$.

Some parameter combinations yield peculiar results, such as extremely high (highly super-Eddington) luminosities,
which will require further investigation; for a few runs the determination of timescales, such
as $t_{3,bol}$ and $t_{m-l}$ may not be accurate, due to fluctuations of the light curve.
For example, the combination $125.30.13$ (standing for 1.25~$\Msun$ WD, 
with $T_{WD}=30\times10^6$~K and $\dot M=10^{-13}\Msun$~yr$^{-1}$) produces
a sharp quick peak at outburst, resulting in very small $t_{3,bol}$ and $t_{m-l}$ (Table~\ref{tbl:outburst}), 
followed by a second increase in luminosity and temperature, but one that does not eject more material. The
values stated in the table relate only to the first quick rise.

\begin{figure}

\plotone{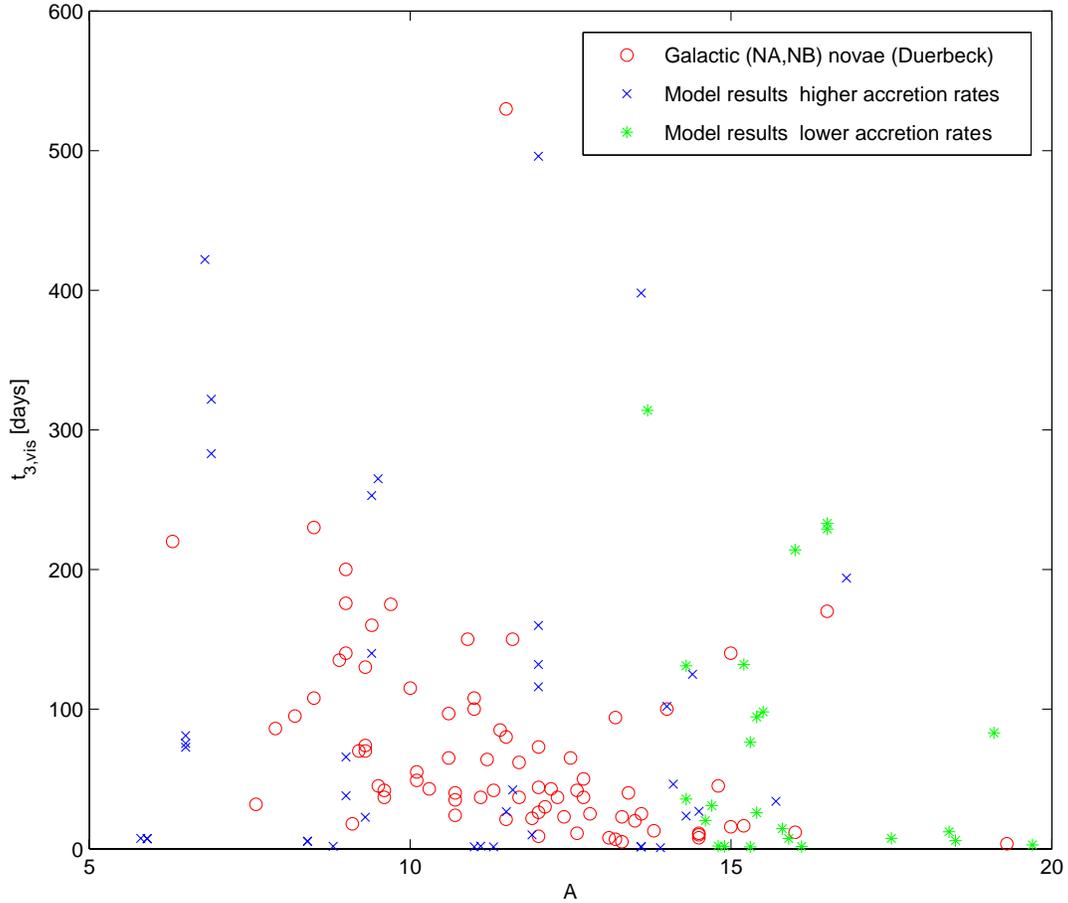} 
\caption{Time of decline $t_{3,vis}$ (order of $t_{m-l}$) 
  versus the outburst amplitude $A$ for
  both observed $NA+NB$ galactic novae (taken from Duerbeck 1987) -
 circles, and our model results: crosses - $\dot{M}\geqq10^{-10}$, asterisks - $\dot{M}\leqq10^{-11}$.}
\label{fig:t3_A_cor} 

\end{figure}

One of the most important characteristics of a classical nova outburst
is the time of decline of the visual magnitude, which serves as a basis for the
"speed class" classification of novae and on which the role of novae as 
standard candles is based.
In Fig.~\ref{fig:t3_A_cor} we show the distribution of the time of
decline $t_{3,vis}$ (which is of the order of $t_{m-l}$ as presented in Table~\ref{tbl:outburst}) 
versus the outburst amplitude $A$ for 79
observed galactic novae from the Duerbeck catalog \citep{Due87},
together with results of the sub-set of 75 models that produced
nova outbursts. Three points should
be noted. First, the general tendency of decreasing $t_3$ for
increasing $A$ is clearly seen. 
Secondly, the model results
provide full coverage of observations in the $[A,t_3]$ plane.
In this respect, it should be noted that the density of points on the plane has no
significance, since the calculated models represent parameter combinations (they are not based on
a population synthesis model).
Finally, we predict the existence of a class of low amplitude novae (with $A\sim7$)
with decline times ranging from 7 days to over a year. We suggest that 
these objects have either not yet been observed, or remain characterized as variables of
some other type.

\begin{figure}

\plotone{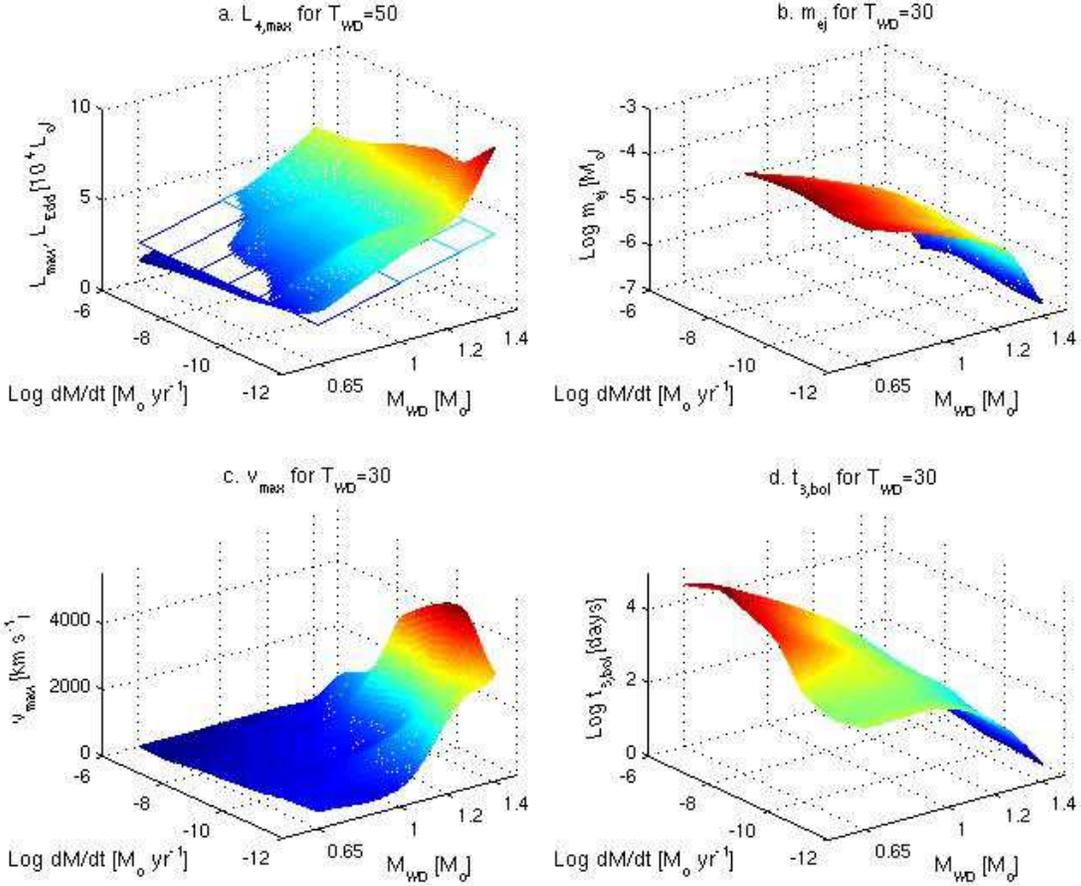}
\caption{3D plots of four major properties out of the complete result
  grid. Each plot represents the property as function of [$\dot{M},
  M_{WD}$] for a representative WD temperature (stated in units of
  $10^6$ K). In panel a, the white surface represents the critical
  Eddington luminosity, emphasizing the domains where the luminosity
  either surpasses or is lower than $L_{Edd}$.}
\label{fig:graphs} 

\end{figure}

The effect of each of the three input parameters on the resulting nova 
characteristic was
already discussed in Paper I and also by \citet{Sch94}. The outcome
is not always self-evident and the trends are not necessarily monotonic. To illustrate
the behavior of some of the more interesting characteristics, we plot
in Fig.~\ref{fig:graphs} the
model results over the $(M_{WD},\dot{M})$ plane for a given $T_{WD}$ 
value. Panel \ref{fig:graphs}a displays the
maximum luminosity attained $L_{4,max}$ together with the $L_{Edd}$
surface, enabling the identification of super- and sub-Eddington
luminosities. The lowest sub-Eddington luminosities are obtained for 
the low $M_{WD}$,
high $\dot M$ corner, identified as the region of least-powerful nova
outbursts. Panel \ref{fig:graphs}b displays the ejected mass $m_{ej}$
(on a logarithmic scale). The highest ejected masses are
obtained for the lowest WD masses through the entire range of accretion
rates. This is because the accreted masses necessary to trigger a TNR are
higher for lower mass WDs (see $\S~3$) whereas at the same time, the 
weaker gravitational
potential enables more massive ejecta, despite the typically lower 
outburst intensity.
In panel \ref{fig:graphs}c we show the maximum
expansion velocities $v_{max}$, which exhibit a similar behavior
to that of the maximum luminosity surface. While the low $M_{WD}$,
high $\dot M$ domain is that of weakest novae, the high $M_{WD}$,
low $\dot M$ corner can be identified as the domain of the
most powerful outbursts.
The time of bolometric decline, as shown in
panel \ref{fig:graphs}d (the trend being very similar to that of $t_{m-l}$, which is not displayed here),
clearly exhibits, in conjunction with figures 2a and 2c,  the correlation between decline time (speed class) and
outburst intensity. The slowest novae (longest decline times) occupy
the domain of weakest outbursts and vice versa.


In Table~\ref{tbl:maxmin} the range of variation of some of the main nova
properties is displayed, according to the new extended grid of
results (for WD-mass range of $0.65-1.00~\Msun$). The parameter combination for which each maximum and minimum
value was obtained is indicated next to the value. We note that
most of the maximum values result from the new, lower $\dot{M}$, runs.


In Table~\ref{obs-calc} we present observed ranges of several nova
characteristics (cf. Prialnik 1995) together with the
ranges obtained from the full grid, for comparison. The
ranges resulting from model calculations overlap and completely
cover the observed ranges.


\section{THE CONSTRAINED PARAMETER SPACE} \label{param_space}

As the extension of the three-dimensional parameter space for novae has resulted in the
emergence of new features, the question arises: where do we stop, if at 
all, extending each one of
the parameters' ranges?
In other words, is the parameter space limited ?
We examine this question based on analytical considerations, verified by
the results of numerical modeling.

\subsection{Heating versus cooling} \label{cool_acc}

\begin{figure}

\plotone{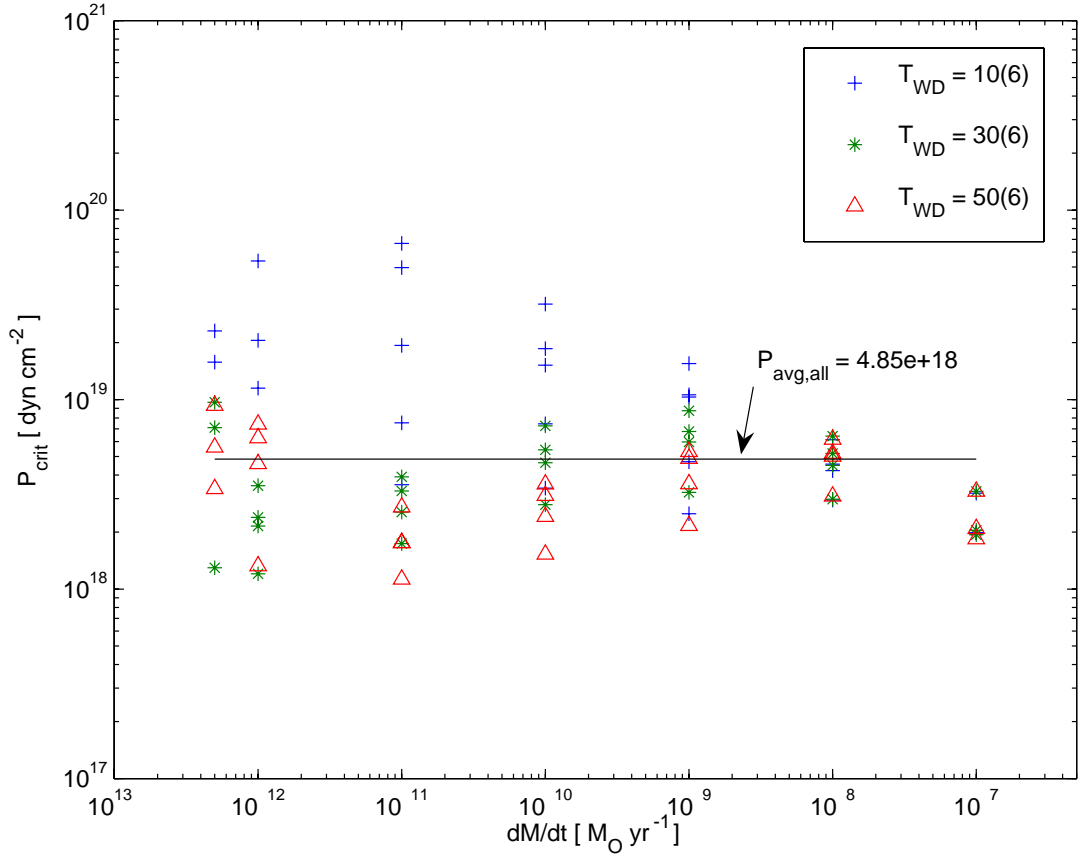}
\caption{Values of the pressure at the base of the envelope prior
  to the TNR, grouped by the three main $T_{WD}$ values as function of the accretion rate; 
  calculated from grid results (according to eq.~\ref{eq:Pcrit}).}
\label{fig:Pcrit} 

\end{figure}

The characteristic timescale for cooling of
a WD, $\tau_{cool}$, is basically a function only of the WD temperature 
\citep{Mes52}.
On the other hand, the accretion timescale $\tau_{acc}$
is directly determined by the accretion rate and indirectly affected by 
the other parameters
through their influence on the mass required to ignite hydrogen,
$\tau_{acc}=m_{acc}(M_{WD},T_{WD},\dot{M})/\dot{M}$. A nova outburst can 
take place only if
$\tau_{cool}>\tau_{acc}$\,; otherwise the temperature cannot rise to the 
point of thermonuclear instability.
A limiting surface is thus obtained by equating these timescales:
\begin{eqnarray}
\label{eq:taucoolsurf}
\tau_{cool}(T_{WD}) - m_{acc}(M_{WD},T_{WD},\dot{M})/\dot{M} & = & 0.
\end{eqnarray}
Analytically, following Mestel (loc. cit.), if we use the Kramers 
opacity law for the WD atmosphere,
the cooling timescale of a CO WD may be approximated by
\begin{equation}
\tau_{cool}\approx2.5\times10^6\left(\frac{M/M_\odot}{L/L_\odot}\right)^{5/7}\ 
{\rm yr},
\end{equation}
while the core temperature is approximated by
\begin{equation}
T_{WD}\approx4\times10^7\left(\frac{L/L_\odot}{M/M_\odot}\right)^{2/7} 
{\rm ~K}.
\end{equation}
On this approximation, then,
\begin{equation}
\tau_{cool}(T_{WD})\approx8\times10^7\left({10^7{\rm K}\over 
T_{WD}}\right)^{2.5}\ {\rm yr}.
\label{ap:taucool}
\end{equation}
An approximation to the accreted mass can be derived from the requirement
that the pressure at the base of the envelope
\begin{equation}
P_b={GM_{WD}m_{acc}\over4\pi R_{WD}^4(M_{WD})}
\label{eq:Pcrit}
\end{equation}
exceed a critical prescribed value, say, $P_{crit}\approx
10^{19}$~dyn~cm$^{-2}$, in order to obtain a TNR 
\citep{Fuj82}. 
Fig.~\ref{fig:Pcrit}, based on grid results, presents yet another
numerical validation of $P_{crit}$ having values of the order of 
$10^{18}-10^{19}
dyn\ cm^{-2}$. We note that at low WD temperatures the range of $P_{crit}$ values
is wider, the critical pressure changing with both $\dot M$ and $M_{WD}$.
Finally, using the Nauenberg approximation \citep{Nau72} to the
Hamada-Salpeter WD mass-radius relation \citep{Ham61},
\begin{equation}
\label{eq:RwdMwd}
{R_{WD}\over R_\odot}\approx1.12\times 
10^{-2}\left[\left(\frac{M_{WD}}{M_{Ch}}\right)^{-2/3}\ -\
\left(\frac{M_{WD}}{M_{Ch}}\right)^{2/3}\right]^{1/2}\ ,
\end{equation}
we obtain $m_{acc}=m_{acc}(M_{WD})$. Substitution of the analytic 
relations into
Eq.~\ref{eq:taucoolsurf} yields the desired surface.

A more accurate result is obtained numerically. First, we calculate 
cooling curves $T_{WD}(t)$ for all the WD masses
of the grid (indeed, they are almost independent of $M_{WD}$). We define 
the cooling timescale at any point by
$\tau_{cool}=(d\ln T/dt)^{-1}$ and calculate it for the $T_{WD}$ values 
of the grid. Then, for each pair
$(T_{WD}, \dot M)$, we solve $m_{acc}(M_{WD}, T_{WD}, \dot M)=\dot 
M\tau_{cool}(T_{WD})$ to obtain $M_{WD}$,
by interpolation on the grid results. We thus obtain the points that 
satisfy Eq.~\ref{eq:taucoolsurf},
defining a surface within the parameter space. It is the bottom surface 
shown in Fig.~\ref{fig:vol-dots}.
The surface exhibits a monotonic decline in $M_{WD}$ 
for each $T_{WD}$, from lower to
higher $\dot{M}$, but also some decline of $M_{WD}$ at the same
accretion rate from higher to lower $T_{WD}$. As mentioned earlier, the
cooling timescales increase with decreasing temperatures. Because longer
$\tau_{cool}$ (relative to $\tau_{acc}$) is what we require for a
nova outburst to develop, 
the WD mass restriction is weaker for lower temperatures.
In principle, if $m_{acc}$ were mainly a function of $M_{WD}$, the condition
for a nova outburst to occur would be eventually satisfied for any WD and
accretion rate, because $\tau_{cool}$ would increase, while $\tau_{acc}$ 
would remain fixed. However, calculations show that as $T_{WD}$ decreases,
$m_{acc}$ becomes ever more strongly dependent on $T_{WD}$, increasing 
steeply as the WD cools. Thus, below a certain value of $T_{WD}$,
the relation $\tau_{cool}\,<\,\tau_{acc}$
is maintained while both timescales increase (cf. \cite{Sch94}).
In other words, when mass accumulates slowly as the WD cools down,
the accreted mass remains below the critical value while both increase with time.

The cooling time constraint is clearly illustrated in 
Fig.~\ref{fig:cooltwo}, where examples are shown
of the evolution of two
characteristics: the maximum temperature within the burning envelope
$T_{max}$ and the total luminosity $L$, for two parameter combinations.
A range of accretion rates is
displayed, following the evolution over a period of one full cycle,
from the beginning of accretion through outburst and decline.
The lowest accretion rate for which a "true" mass-ejecting nova is
obtained is $\dot{M}=5\times10^{-13}~M_\odot$~yr$^{-1}$ (and even then,
not for all parameter combinations, as seen in the results
grid). We see in Fig.~\ref{fig:cooltwo} that for a slightly lower
accretion rate value $2.5\times10^{-13}~M_\odot$~yr$^{-1}$, the WD cools
down faster than it is able to accrete sufficient mass to reach TNR
conditions. This is clearly seen in the
$T_{max}(t)$ plot: the curve gradually declines, without having reached
an outburst. Equally, the luminosity curve remains flat for the lowest
$\dot{M}$. The curves for the higher accretion
rates exhibit outburst events where $T_{max}(t)$ exceeds $10^8$~K
and the bolometric luminosity increases above $4-5\times10^4~L_\odot$.
Note also the obvious but nicely
demonstrated fact that as we move from lower to higher accretion
rates, the outburst occurs earlier on the timeline, hence the recurrence
period becomes shorter.

\subsection{Nuclear versus gravitational energy} \label{nuc_grav}

The source of energy for classical nova outbursts is nuclear energy 
released during the TNR, by
burning a fraction $f$ of the hydrogen content of the accreted mass. Thus
\begin{equation}
\label{eq:Enuc}
E_{nuc}=fXm_{acc}Q,
\end{equation}
where: $X\approx0.7$ is the hydrogen mass fraction in the outer layers 
of the nova companion star, and
$Q\approx6\times10^{18}$~erg~g$^{-1}$ is the energy released per gram of 
burnt hydrogen.

The greatest part of the energy released at outburst is used in lifting 
the ejected
shell from the gravitational potential well of the white dwarf. This 
gravitational energy
may be roughly approximated by
\begin{equation}
\label{eq:Egrav}
E_{grav}=\frac{GM_{WD}m_{ej}}{R_{WD}}.
\end{equation}
Obviously, a mass ejecting outburst can take place only if 
$E_{nuc}>E_{grav}$. Therefore,
a limiting surface may be defined by requiring $E_{nuc}=E_{grav}$,
\begin{equation}
\label{eq:EgravEnucSurf}
{m_{acc}(M_{WD},\dot{M},T_{WD})\over m_{ej}(M_{WD},\dot{M},T_{WD})}\ -\ 
\frac{G}{fXQ}\frac{M_{WD}}{R_{WD}}\,=\,0.
\end{equation}
As a rough analytical estimate, the difference between $m_{acc}$ and 
$m_{ej}$, may be neglected,
in which case Eq.(\ref{eq:EgravEnucSurf}) simply
imposes an upper limit on $M_{WD}$ for a given value of $f$ (e.g., 
1.22~$M_\odot$ for $f=0.1$ and 1.42~$M_\odot$ for $f=0.3$),
a flat surface parallel to the [$T_{WD},\dot M$] plane in the parameter
space. This is already a more severe constraint than just the 
Chandrasekhar mass limit.
Using the results of the numerical computations and taking $f=0.3$ for 
illustration, we obtain a more significant constraint in the
form of a slightly curved surface, the top surface in Fig.~\ref{fig:vol-dots}.

\subsection{Accretion versus Eddington luminosity} \label{acc_edd}

For the nova progenitor to be able to accrete material
during the quiescence phase, the accretion luminosity $L_{acc}$
must be lower than the Eddington critical luminosity $L_{Edd}$.
Otherwise, radiation
pressure would push away and dissipate the accreted
material. In fact, the total (net) luminosity of the accreting star 
should be lower
than the Eddington limit, but in most cases the WD intrinsic luminosity
is negligible compared with $L_{acc}$. The accretion luminosity is given by
\begin{equation}
\label{eq:Lacc}
L_{acc}=\alpha GM_{WD}\dot{M}/R_{WD},
\end{equation}
where $\alpha\sim0.15$, taking into account accretional heating 
\citep{Reg89},
and again, using Eq.~\ref{eq:RwdMwd} for $R_{WD}(M_{WD})$.
Thus a third limiting surface, obtained by equating $L_{acc}$ and $L_{Edd}$,
\begin{equation}
\label{eq:LaccLeddSurf}
\dot{M}\ -\ \frac{4\pi c}{\alpha\kappa_s}R_{WD}(M_{WD})\,=\,0
\end{equation}
(where $\kappa_s$ is the electron scattering opacity coefficient) 
is shown in Fig.~\ref{fig:vol-dots}.
We note that the result, which in this case relies solely on analytical 
considerations, is independent of
the WD temperature, although it might be indirectly affected by it to 
some extent through a more realistic opacity coefficient.
It is not surprising that the allowed accretion
rates decrease with increasing $M_{WD}$.

\subsection{Degeneracy and the WD core temperature} \label{wdtemperature}

\begin{figure}

\plotone{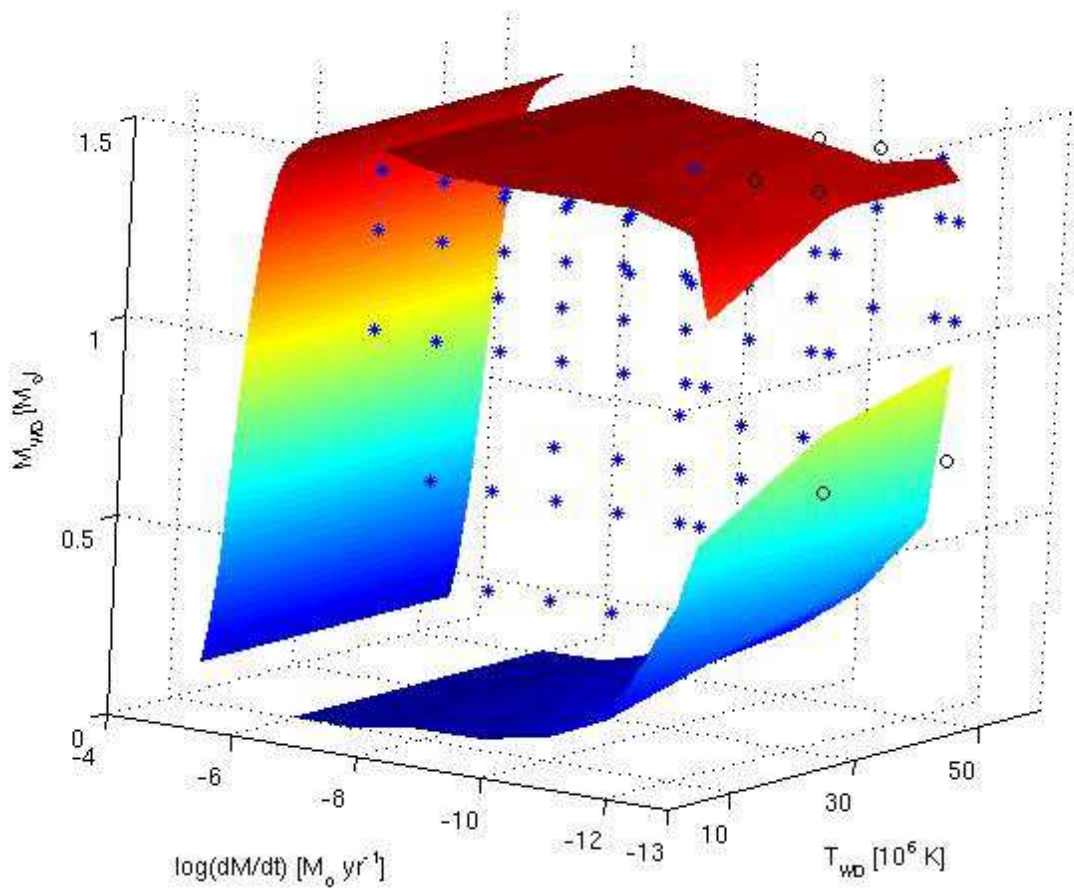}
\caption{The combined restricting surfaces, confining a volume within the 3D parameter
  space where conditions for nova outbursts are satisfied. The bottom
  surface corresponds to the heating versus cooling criterion (section
  3.1); the top WD mass-limiting surface corresponds to the nuclear
  versus gravitational energy considerations (section 3.2); the left
  mass transfer rate-limiting surface relates to the accretion versus
  Eddington luminosity (section 3.3). Positions of the grid parameter
  combinations that successfully produced nova eruptions are plotted within the confined volume.}
\label{fig:vol-dots} 

\end{figure}

\begin{figure}

\plottwo{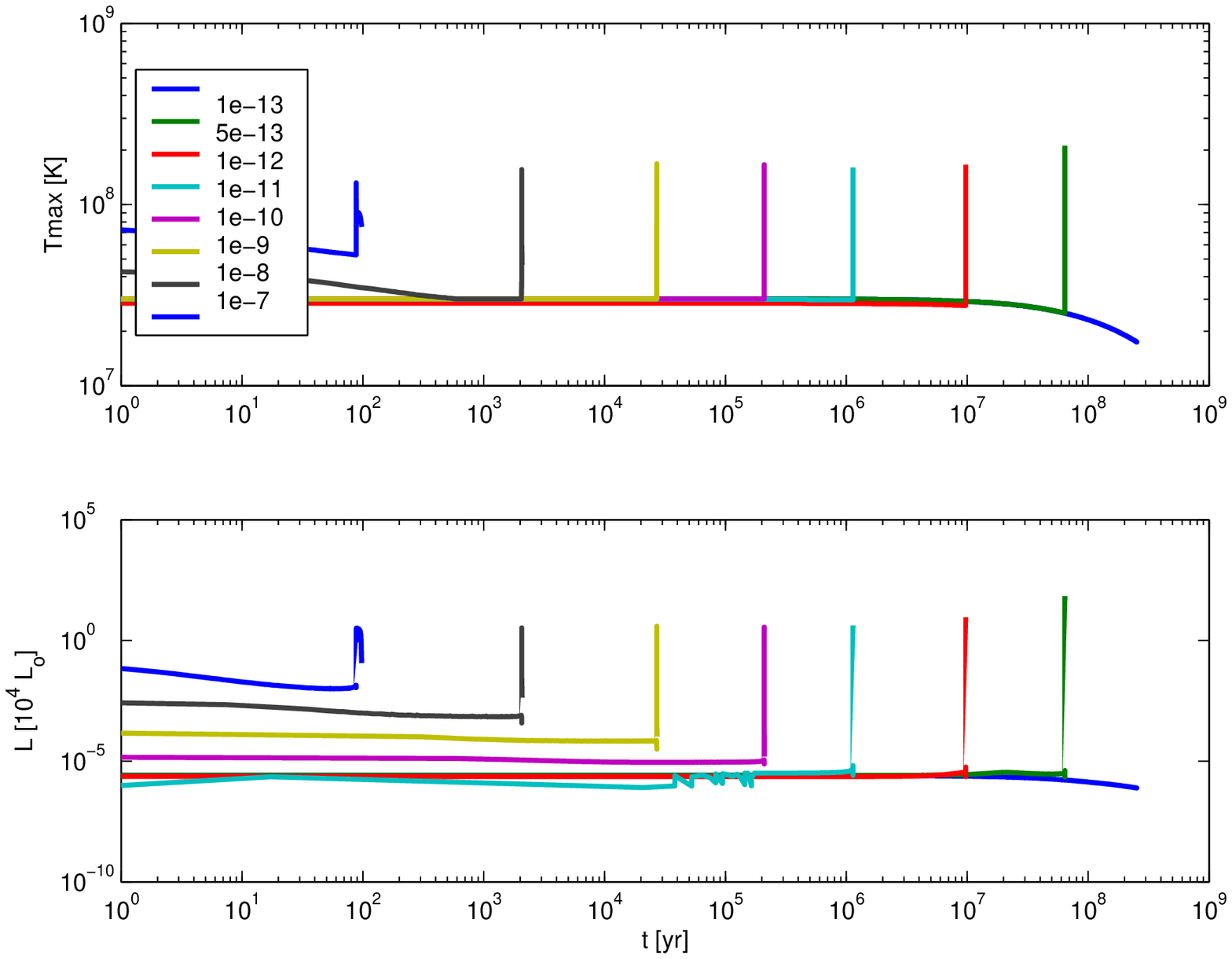}{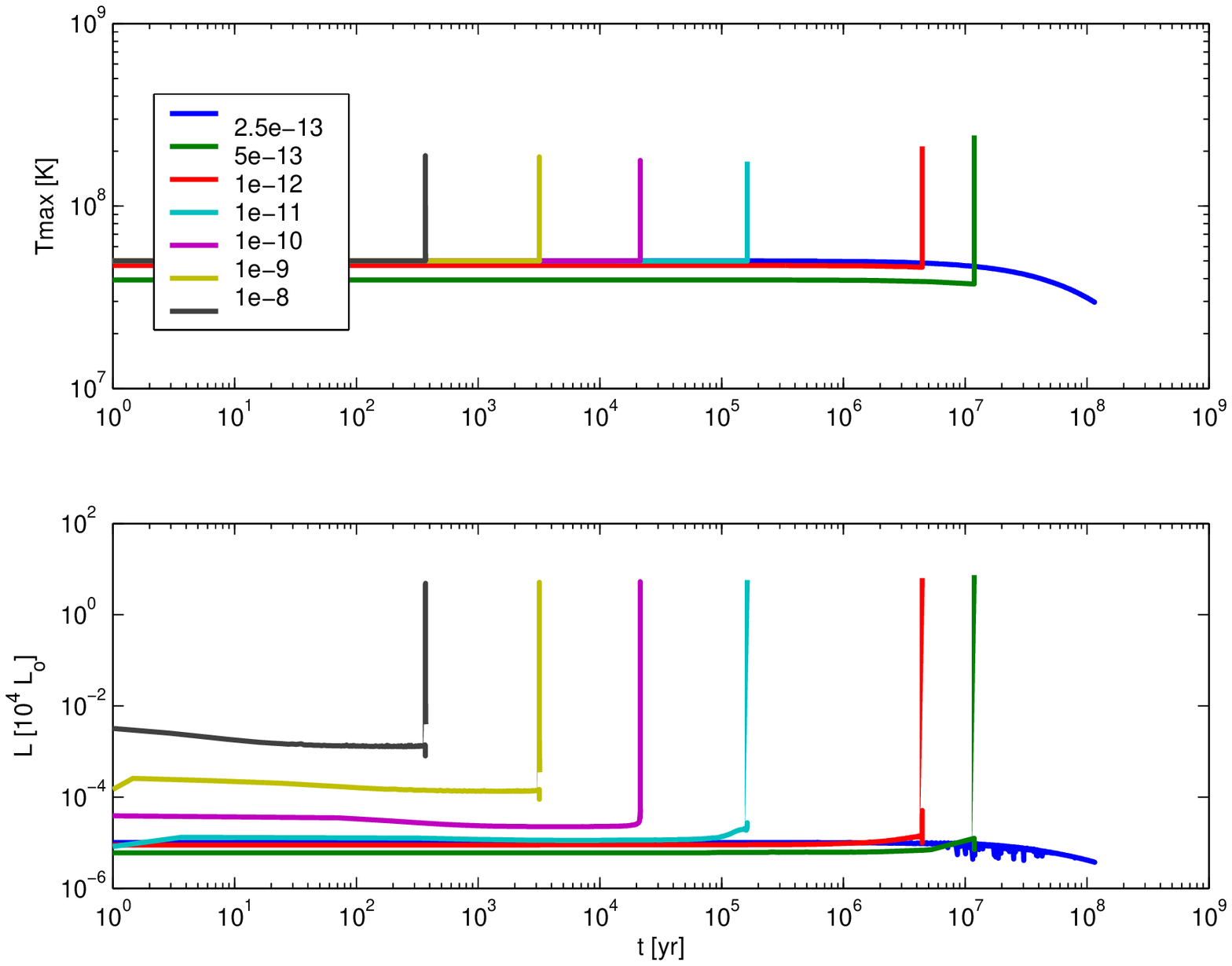}
\caption{Evolution in time of two characteristics for different accretion rates: 
  the maximum temperature (within the burning shells)
  $T_{max}$ and the luminosity $L$, for $M_{WD}=1.00 M_{\odot},
  T_{WD}=30\times10^6 K$ (left), and $M_{WD}=1.25 M_{\odot},
  T_{WD}=50\times10^6 K$ (right).}
\label{fig:cooltwo} 

\end{figure}

According to Fig.~\ref{fig:vol-dots}, the relevant region in parameter space
where nova eruptions can occur is tube-shaped. Our calculations assumed that the
relevant range of WD core temperatures is $10-50\times10^6$~K, taking
the discrete representative values of 10, 30, and 50$\times10^6$~K. 
In fact, the WD temperature is restricted from above and below, as we shall now show.

We may obtain a rough estimate for the upper limit on $T_{WD}$
by setting the Fermi parameter $\varepsilon_f\propto(\ln{P}-2.5\ln{T})$ to
zero at the base of the accreted layer, where the pressure, 
approximately given by
Eq.~\ref{eq:Pcrit}, is of the order of $10^{19}$~dyn~cm$^{-2}$.
Thus assuming that $T_{WD}\approx T_b$, we have
\begin{equation}
T_{WD}\approx4.7\times10^7\left({P_e/10^{19}\,{\rm dyn~cm}^{-2}\over 
{\rm e}^{\varepsilon_f}}\right)^{2/5}\ {\rm K},
\label{eq:TfermiSurf}
\end{equation}
where $P_e$ is the electron pressure. Hence if the condition of strong
electron degeneracy is expressed as $\varepsilon_f\apgt0$, it results in 
the requirement $T_{WD}\aplt5\times10^7$~K. Therefore the tube of Fig.~\ref{fig:vol-dots}
has a high temperature end at a $T_{WD}$ of about 50 million degrees.
Does it have a low temperature end?

A lower limit on $T_{WD}$ results from the following considerations.
When material starts accumulating on the WD surface, its
temperature is lower than the core temperature and also lower than the ignition
temperature (otherwise hydrogen would ignite immediately and quietly, rather
than explosively under degenerate conditions).
As the material becomes compressed, it releases gravitational energy, which is absorbed, in part, by
the accreted layer, while in part it is conducted into the core.
If the absorption of heat is sufficiently effective to raise the temperature of the hydrogen-rich
material then, eventually, the temperature at the bottom of the hydrogen-rich layer will 
become high enough for hydrogen to ignite. The nuclear luminosity, low at first, will soon become
the dominant energy source. Therefore, the  restrictive condition for a nova outburst to occur is
that compressional heating be sufficiently effective in order to raise the temperature
to the ignition value of roughly $15\times10^6$~K required by the CNO cycle. The competition between
compressional heating power and the rate of heat conduction can only be decided by
solving the energy equation. Although it is reasonable to assume that accumulation of matter at a high rate
and/or a massive WD will tip the scale in favor of heating,
an accurate analytical result is difficult to obtain; 
it would require too many simplifying assumptions and approximations.
We therefore resort to numerical modeling
in order to obtain an estimate on the limiting temperature. 

A few additional models, beyond the grid described in Section \ref{numerical} were calculated for WDs cooler than
$10^7$~K; indeed very old WDs in binary systems may reach temperatures below $10^7$~K \citep{Nel04}.
For high accretion
rates ($\dot M=10^{-9}\Msun$~yr$^{-1}$), we obtained nova outbursts even for 
$T_{WD}=3\times10^6$~K and the results were very similar to those obtained for 
$T_{WD}=10^7$~K. At the other $\dot M$ end, adopting $\dot M=10^{-12}\Msun$~yr$^{-1}$ and $M_{WD}=1~\Msun$,
we did {\it not} obtain an outburst for $T_{WD}/10^6$~K=5, 7 and 8; the accreted material 
cooled continuously. However, for $T_{WD}=9\times10^6$~K a regular nova outburst was obtained.
Thus the lower limit for $T_{WD}$ is strongly dependent on $\dot M$, as expected, and 
in agreement with the conclusions of \citet{Sch94}. 
In the case of a very cold WD, heating is impeded by heat conduction into the hydrogen-depleted
deep core and may be altogether suppressed. However, the impediment is less severe at high 
accretion rates, due to the rapid supply of accretion energy.

We also checked the effect of $M_{WD}$:
taking $T_{WD}/10^6$~K=8 (no nova for $M_{WD}=1~\Msun$), but
increasing the WD mass to 1.25$\Msun$, we did obtain a nova outburst. In this case we found 
the lower limit to be below $8\times10^6$~K but above $7\times10^6$~K.
Thus, a lower limit to the WD core temperature exists for any combination of WD mass
and accretion rate, but for high accretion rates this limit is so low that it requires more
than the age of the universe for a cooling WD to reach it. Hence the constraint on $T_{WD}$
becomes significant only in the case of low accretion rates.

Finally, since it is not clear that erupting WDs may cool down to the theoretical lower limit,
and since between the lower limit and $10^7$~K
the results are not highly sensitive to the precise value
of $T_{WD}$ (cf. \citet{Sch94}), we chose $10^7$~K
as the lowest $T_{WD}$ value in the present study, as in Paper I. 
However, accretion on very cold WDs will be further investigated in a future paper, since it is
this part of the parameter space that appears likely to shed light on the most peculiar observed
novae, showing very high luminosities, relatively low expansion velocities and, apparently,
large ejected masses.

\subsection{The full parameter space} \label{fullspace}

The three surfaces obtained so far and displayed in 
Fig.~\ref{fig:vol-dots} describe a restricted, tube-shaped
region within the 3D parameter space where
conditions for nova outbursts are expected to be satisfied. The
additional constraints imposed by the electron degeneracy and mainly
concerning the WD temperature determine the ends of the tube. 

Hence there appears to be a confined volume of parameter space, 
where conditions for classical nova outbursts are satisfied.
Plotted on top of these surfaces are all of the parameter combination positions that
produced "well-behaved" mass-ejecting nova outbursts in our runs (including the three runs for
the lowest WD mass of $0.40 \Msun$ and two runs for WD temperatures lower than $10^7~K$ as analyzed
in the previous section).
There are six "mass ejecting" parameter combinations that lie on the 
boundaries of the restricting volume, denoted in Fig.~\ref{fig:vol-dots} by black circles, while the other
75 combinations lie well inside the volume, and are denoted by 
blue asterisks. As we have explained earlier, the confining volume we have 
constructed emphasizes the existence of limitations (on each parameter range) and
also a fairly clear shape, but one that should be understood as 
qualitative. Positions of mass-ejecting nova parameter combinations of
the numerical calculations may lie on the verge or even slightly
outside that volume, since construction of the different surfaces
involves many constants, coefficients and approximations, influencing
the exact values obtained. It would have been very instructive were we
also able to place on this graph the positions of (parameter 
combinations of)
observed novae. Unfortunately, the estimation of all three parameters
for any observed eruption is still problematic and involves too many
uncertainties. A unique estimation of the three basic
parameters that lead to a set of observed characteristics is an
ambitious aim yet to be achieved; hopefully it will be, with the help
of studies such as the present one.

\section{SUMMARY AND CONCLUSIONS} \label{conclusions}

In the framework of this study we have extended the 3D parameter grid
of multicycle nova evolution to cover combinations with low accretion
rates. The main results of this study may be summarized as follows:

1. The entire range of observed nova characteristics is
thoroughly covered by the complete grid of models. Even exceptional
observed values such as outburst amplitude of over 19 magnitudes and 
very high $Z$ are
covered by the new grid of results. The majority of the calculated
maximum values for the various characteristics is obtained for the
lower-$\dot{M}$ runs, down to the lowest accretion rate value of
$5\times10^{-13}~M_\odot$~yr$^{-1}$.

2. There has been a problem deriving ejected mass values as high as
those deduced observationally. Moreover, a strong case has been made that
much of the ejecta from each nova is missed by observers \citep{Fer98}. We have managed to
produce a notably high $m_{ej}$ value of $6.6\times10^{-4}~M_\odot$,
which is almost a factor $3$ greater than the maximum value
obtained in the original grid (Paper I). This high  $m_{ej}$ was produced
for a low $M_{WD}$ at low $\dot{M}$ (combination $065.10.12$). The
former maximum  $m_{ej}$ was obtained for the same $[M_{WD},T_{WD}]$
combination but with $\dot{M}$ of $10^{-10}~M_\odot$~yr$^{-1}$.

3. For the new runs of lower  $\dot{M}$ values we obtain maximum
outburst luminosities surpassing the Eddington luminosities
(calculated for electron scattering opacities) by factors of up to a
few tens, for the whole range of WD masses. The highest
super-Eddington luminosities (with correspondingly highest derived
outburst amplitudes) are obtained for fast to very-fast novae.

4. We predict the existence of remarkably small amplitude novae with decline times 
ranging from a week to over a year.

5. In previous studies, as well as the current one, we have seen
that by extending the parameter space new features emerge. That raised
the question whether the parameter space is at all limited, and if so,
what are the exact limits on the three basic independent
parameters? 
Using both models and analytic relations, we derived a
confining volume within the three dimensional parameter space, where conditions for
nova outbursts are satisfied.
Thus the parameter space is indeed
limited. 
As the new grid of models now covers the entire parameter space (assuming WDs composed of
CO --- but see below),
it should enable us to provide explanations for most (all?)
novae, including the more peculiar objects among them. This, however, will be
the subject of a later study.
Of special significance is the existence
of a critical accretion rate $\dot{M}_{crit}(M_{WD},T_{WD})$ below
which conditions are not sufficient to trigger a TNR. The critical
mass transfer rate is indeed very low, but seems to be relevant for old
close interacting binaries with very low mass companions.
For cases like these, the determination of limits is important.

6. Although this study covers the entire range of WD masses, practically
up to the Chandrasekhar limit, the assumed composition of the WD core
is C and O (in equal mass fractions) for all models. At first glance, this might appear
unrealistic; very massive WDs are believed to
emerge after the carbon burning stage in the evolution of a star, and 
they should be composed of oxygen, neon and magnesium \citep{Gut96}.
Although the mass fractions of these elements and their dependence on
the WD mass are still uncertain issues, clearly such WDs should not
contain carbon. On the other hand, our grid of models produces results that
cover the {\it entire}
range of observed characteristics (ignoring the breakdown of the total 
heavy element
mass fraction $Z_{ej}$). This seems to indicate that they are more widely
applicable than imposed by the assumption on the WD composition.
In fact, the mechanism of nova eruptions revolves around thermonuclear instability under
conditions of electron degeneracy. Electron degeneracy, in turn, is 
determined
by $\mu_e=<A/Z>$, which assumes the same value for helium and carbon 
burning products.
In addition, the CNO cycle is not sensitive to the initial CNO breakdown. 
Consequently, we should not expect significant differences in the outburst 
characteristics of WDs that differ only in composition, except for the {\it breakdown} of 
$Z_{ej}$. In order to test this expectation we have repeated the calculations for two
illustrative models after replacing the carbon in the WD core by neon. 
The results are summarized in Table~\ref{tbl:CO-ONe} and indeed they confirm our 
prediction. Thus it is not surprising that the entire range of nova characteristics is
reproduced by our CO models and the results should be applicable to 
novae in general.

\begin{figure}
\epsscale{.60}
\plotone{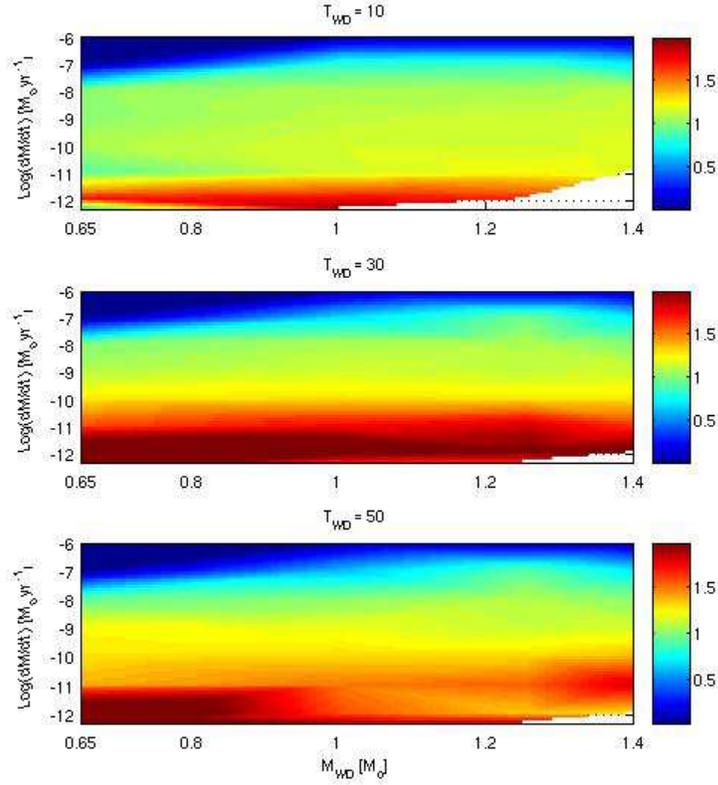}
\caption{Color maps displaying shades that correspond to values of the ratio $m_{ej}/m_{acc}$
  in the $[M_{WD},Log\dot M]$ plane. A value of 1 corresponds to an unevolving WD mass,
  below - an increase in $M_{WD}$, and above - a decrease in time of $M_{WD}$.
  The three panels correspond to the three $T_{WD}$ values - 10, 30 and 50 $(10^6 K)$, top to bottom.}

\label{fig:macc_ej} 

\end{figure}

\begin{figure}
\epsscale{1.}
\plotone{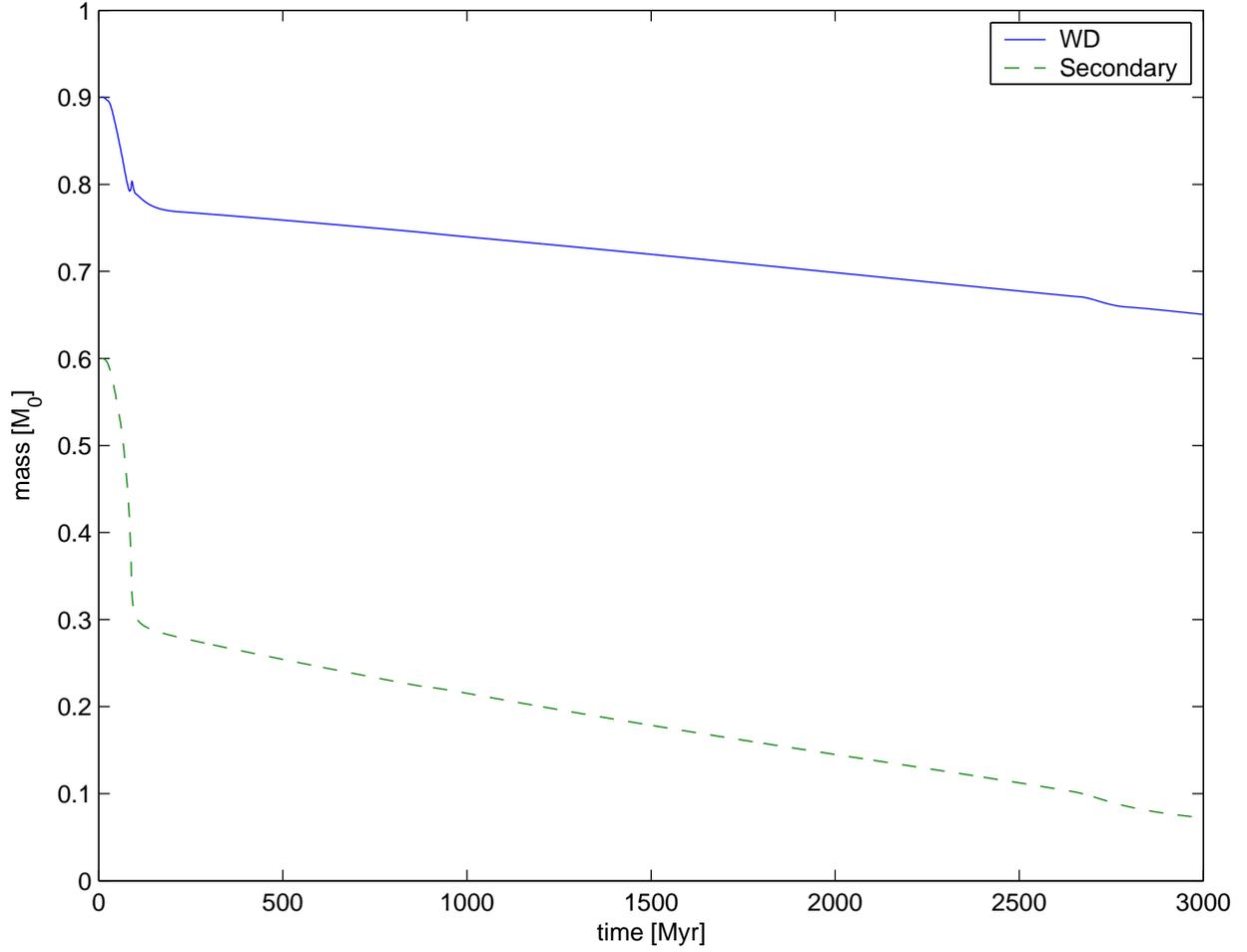}
\caption{An example calculation of $M_{WD}(t)$ based on an evolving
  $\dot M$ for initial masses of $0.6$ and $0.9\ M_\odot$ of the
  secondary and primary respectively.}
\label{fig:wdmass} 

\end{figure}

7. An intriguing issue related to the investigation of novae is the
question of the ultimate fate of the WD. Will it be losing or
gaining mass after multiple successive cycles? Will it be able 
to reach $M_{Ch}$, thus acting as a possible SNIa precursor, or not?
The ratio $m_{ej}/m_{acc}$ is shown in Fig.~\ref{fig:macc_ej} for all parameter combinations.  
We note that it falls below unity only in a small region of the parameter space,
thus strongly reducing the possibility of SNIa to result from accreting WDs.
Nevertheless, this region does lead continuously from low mass to the Chandrasekhar limit
provided the accretion rate remains very high all along. Therefore, it is possible, at least
in principle, for a WD to grow by accretion up to $M_{Ch}$. In order to further investigate
this problem we need to consider the dynamic evolution of the binary system that determines
the evolution of mass transfer rate.
Given the accretion rate as a function
of time $\dot M_{acc}(t)$, which changes with binary separation and 
masses, we can calculate the change of $M_{WD}$ with time,
\begin{equation}
M_{WD}(t)\ =\ \int [\dot M_{acc}(t)\ -\ \dot M_{ej}(\dot 
M_{acc}(t),M_{WD}(t))]\,dt,
\end{equation}
where
\begin{equation}
\dot M_{ej}(\dot M_{acc}(t),M_{WD}(t))\ =\ m_{ej}/P_{rec}
\end{equation}
is obtained by interpolation on the grid. An example is shown in 
Fig.~\ref{fig:wdmass}, based on
an evolving accretion rate kindly supplied by Jarrod Hurley, for a 
binary system of
initial masses 0.6~$M_\odot$ (secondary) and 0.9~$M_\odot$ (primary).

8. Finally, in a similar
manner, the evolving luminosity of an accreting WD may be constructed as 
a quasi-periodic
function of the variables $L_{acc}$, $L_{max}$, $t_{3,bol}$ and 
$P_{rec}$ supplied by the grid,
all of which change with time as $M_{WD}$ and $\dot M_{acc}$ change.
Thus, in future, an important use of the nova grid will be the
parameterization of nova evolution for long term modeling of large 
stellar systems.

\acknowledgements
We thank Jarrod Hurley for providing us with his results on the genesis 
of low-mass white dwarfs in close binaries (the models of 
Table~\ref{tbl:bin-evol} and Fig.~\ref{fig:wdmass}). We also thank an
anonymous referee for valuable comments and suggestions.

\begin{deluxetable}{rccrrrccl}

\tablecaption{Example of Binary Evolution Sequence - Creation of a Low-Mass CO WD\label{tbl:bin-evol}}
\tablewidth{0pt}
\tablehead{
\colhead{Time (Myr)} & \colhead{M1} & \colhead{M2} &
\multicolumn{2}{c}{Types\tablenotemark{*}} &
\colhead{Sep. (AU)} & \colhead{R1/RL1} & \colhead{R2/RL2} &
\colhead{}
}
\startdata
$0.000$     & $2.809$ & $0.196$ &   MS & MS & $110.363$ & $0.03$ &  $0.01$ & initial\\
$450.482$   & $2.809$ & $0.196$ &   HG & MS & $110.363$ & $0.07$ &  $0.01$ & ev change\\
$453.303$   & $2.809$ & $0.196$ &   GB & MS & $110.408$ & $0.19$ &  $0.01$ & ev change\\
$456.049$   & $2.808$ & $0.196$ &   GB & MS & $49.259$  & $1.01$ &  $0.03$ & begin RLOF\\
$456.049$   & $0.414$ & $0.196$ & HeMS & MS &  $0.854$  & $1.01$ &  $0.03$ & common env\\
$758.047$   & $0.413$ & $0.196$ & HeGB & MS &  $0.782$  & $0.25$ &  $0.92$ & ev change\\
$786.929$   & $0.412$ & $0.196$ & COWD & MS &  $0.765$  & $0.05$ &  $0.94$ & ev change\\
$1010.000$  & $0.412$ & $0.196$ & COWD & MS &  $0.718$  & $0.05$ &  $1.01$ & begin RLOF\tablenotemark{**}\\
$12000.000$ & $0.412$ & $0.040$ & COWD & MS &  $0.841$  & $0.03$ &  $1.06$ & max time\\
\enddata

\tablenotetext{*}{MS = Main Sequence, HG = Hertzsprung Gap, GB = Giant Branch.}
\tablenotetext{**}{From this stage onwards CV behavior - continuous mass-loss from secondary to binary 
with periodic mass ejections.(RLOF = Roche-lobe overflow)}

\end{deluxetable}

\begin{deluxetable}{ccc|ccccccc}

\tablecaption{Complete Grid Results - Characteristics of The Nova Envelope\label{tbl:envelope}}
\tabletypesize{\scriptsize}
\tablewidth{0pt}
\tablehead{
\multicolumn{3}{c}{Param. Comb.} &
\multicolumn{7}{c}{Nova Envelope Characteristics} \\
\colhead{$M_{WD}$} & \colhead{$T_{WD}$} & \colhead{$log\dot{M}$} &
\colhead{$m_{acc}$} & \colhead{$m_{ej}$} & \colhead{$Y_{env}$} &
\colhead{$Y_{ej}$} & \colhead{$Z_{env}$} & \colhead{$Z_{ej}$} & \colhead{$T_{8,max}$}\\
\colhead{$(M_\odot)$} & \colhead{$(10^6~K)$} &
\colhead{$(M_\odot~yr^{-1})$} & \colhead{$(M_\odot)$} &
\colhead{$(M_\odot)$} & \colhead{} & \colhead{} & \colhead{} & \colhead{} & \colhead{$(10^8~\degr K)$} 
}
\startdata
$0.40$ & $10$ &  $-9$   & $4.12E-04$ & $4.55E-04$ & $0.2568$ & $0.2579$ & $0.1154$ & $0.1204$ & $0.90$\\
$0.40$ & $10$ & $-10$   & $5.62E-04$ & $6.76E-04$ & $0.2358$ & $0.2364$ & $0.1832$ & $0.1893$ & $0.98$\\
$0.40$ & $10$ & $-11$   & $5.87E-04$ & $6.96E-04$ & $0.2444$ & $0.2322$ & $0.1539$ & $0.2075$ & $0.95$\\
\hline
$0.65$ & $10$ &  $-6$   & $8.35E-06$ & $0.00E+00$ & $0.2830$ & {\nodata}& $0.0208$ & {\nodata}& $0.81$\\
$0.65$ & $10$ &  $-7$   & $2.45E-05$ & $0.00E+00$ & $0.3434$ & {\nodata}& $0.0203$ & {\nodata}& $1.10$\\
$0.65$ & $10$ &  $-8$   & $1.01E-04$ & $1.03E-04$ & $0.3754$ & $0.3865$ & $0.0207$ & $0.0215$ & $1.34$\\
$0.65$ & $10$ &  $-9$   & $1.61E-04$ & $1.63E-04$ & $0.2540$ & $0.2650$ & $0.1201$ & $0.1350$ & $1.39$\\
$0.65$ & $10$ & $-10$   & $2.55E-04$ & $2.76E-04$ & $0.2407$ & $0.2498$ & $0.1637$ & $0.1786$ & $1.67$\\
$0.65$ & $10$ & $-11$   & $2.58E-04$ & $2.37E-04$ & $0.2570$ & $0.2538$ & $0.1233$ & $0.1847$ & $1.39$\\
$0.65$ & $10$ & $-12$   & $3.94E-04$ & $6.66E-04$ & $0.2660$ & $0.1799$ & $0.0500$ & $0.4458$ & $1.59$\\
$0.65$ & $10$ & $-12.3$\tablenotemark{*}  & $5.40E-04$ & $5.04E-04$ & $0.2644$ & $0.2723$ & $0.0504$ & $0.0733$ & $1.51$\\
\hline
$0.65$ & $30$ &  $-6$   & $8.63E-06$ & $0.00E+00$ & $0.2830$ & {\nodata}& $0.0208$ & {\nodata}& $0.80$\\
$0.65$ & $30$ &  $-7$   & $2.54E-05$ & $0.00E+00$ & $0.3690$ & {\nodata}& $0.0203$ & {\nodata}& $1.09$\\
$0.65$ & $30$ &  $-8$   & $1.02E-04$ & $1.01E-04$ & $0.3689$ & $0.3801$ & $0.0208$ & $0.0214$ & $1.22$\\
$0.65$ & $30$ &  $-9$   & $1.11E-04$ & $1.21E-04$ & $0.2884$ & $0.2956$ & $0.1037$ & $0.1102$ & $1.21$\\
$0.65$ & $30$ & $-10$   & $9.55E-05$ & $1.21E-04$ & $0.2462$ & $0.2489$ & $0.2247$ & $0.2370$ & $1.24$\\
$0.65$ & $30$ & $-11$   & $5.96E-05$ & $9.46E-05$ & $0.1804$ & $0.1817$ & $0.3767$ & $0.3910$ & $1.21$\\
$0.65$ & $30$ & $-12$   & $4.12E-05$ & $9.61E-05$ & $0.1203$ & $0.1216$ & $0.5961$ & $0.6085$ & $1.28$\\
$0.65$ & $30$ & $-12.3$ & $4.44E-05$ & $1.07E-04$ & $0.1149$ & $0.1161$ & $0.6170$ & $0.6300$ & $1.32$\\
\hline
$0.65$ & $50$ &  $-6$   & $8.94E-06$ & $0.00E+00$ & $0.2860$ & {\nodata}& $0.0206$ & {\nodata}& $0.80$\\
$0.65$ & $50$ &  $-7$   & $2.66E-05$ & $0.00E+00$ & $0.3497$ & {\nodata}& $0.0203$ & {\nodata}& $0.97$\\
$0.65$ & $50$ &  $-8$   & $1.06E-04$ & $9.88E-05$ & $0.3669$ & $0.3774$ & $0.0207$ & $0.0212$ & $1.31$\\
$0.65$ & $50$ &  $-9$   & $7.41E-05$ & $9.16E-05$ & $0.2556$ & $0.2584$ & $0.2098$ & $0.2208$ & $1.16$\\
$0.65$ & $50$ & $-10$   & $5.23E-05$ & $6.72E-05$ & $0.2549$ & $0.2570$ & $0.2442$ & $0.2551$ & $1.20$\\
$0.65$ & $50$ & $-11$   & $3.86E-05$ & $5.36E-05$ & $0.2480$ & $0.2550$ & $0.3100$ & $0.3160$ & $1.09$\\
$0.65$ & $50$ & $-12$   & $4.54E-05$ & $1.11E-04$ & $0.1320$ & $0.1347$ & $0.6049$ & $0.6171$ & $1.33$\\
$0.65$ & $50$ & $-12.3$ & $1.16E-04$ & $2.61E-04$ & $0.1288$ & $0.1472$ & $0.6262$ & $0.6191$ & $1.60$\\
\hline
$1.00$ & $10$ &  $-6$   & $2.05E-06$ & $0.00E+00$ & $0.3260$ & {\nodata}& $0.0204$ & {\nodata}& $1.05$\\
$1.00$ & $10$ &  $-7$   & $8.96E-06$ & $7.21E-06$ & $0.3421$ & $0.4187$ & $0.0210$ & $0.0223$ & $1.41$\\
$1.00$ & $10$ &  $-8$   & $2.06E-05$ & $2.22E-05$ & $0.3044$ & $0.3314$ & $0.0937$ & $0.1002$ & $1.59$\\
$1.00$ & $10$ &  $-9$   & $4.66E-05$ & $5.18E-05$ & $0.2676$ & $0.2969$ & $0.1200$ & $0.1270$ & $1.86$\\
$1.00$ & $10$ & $-10$   & $8.40E-05$ & $9.72E-05$ & $0.2427$ & $0.2744$ & $0.1553$ & $0.1620$ & $2.10$\\
$1.00$ & $10$ & $-11$   & $8.72E-05$ & $1.00E-04$ & $0.2475$ & $0.2730$ & $0.1529$ & $0.1917$ & $2.09$\\
$1.00$ & $10$ & $-12$   & $9.28E-05$ & $1.62E-04$ & $0.2429$ & $0.1901$ & $0.0332$ & $0.4770$ & $2.10$\\
$1.00$ & $10$ & $-12.3$ & $1.04E-04$ & $2.00E-04$ & $0.1860$ & $0.1772$ & $0.0136$ & $0.5328$ & $2.31$\\
\hline
$1.00$ & $30$ &  $-6$   & $2.10E-06$ & $0.00E+00$ & $0.3160$ & {\nodata}& $0.0204$ & {\nodata}& $1.04$\\
$1.00$ & $30$ &  $-7$   & $8.74E-06$ & $5.74E-06$ & $0.3280$ & $0.3535$ & $0.0208$ & $0.0215$ & $1.34$\\
$1.00$ & $30$ &  $-8$   & $2.03E-05$ & $2.19E-05$ & $0.3035$ & $0.3300$ & $0.0951$ & $0.1010$ & $1.57$\\
$1.00$ & $30$ &  $-9$   & $2.70E-05$ & $3.10E-05$ & $0.2686$ & $0.2940$ & $0.1492$ & $0.1590$ & $1.68$\\
$1.00$ & $30$ & $-10$   & $2.10E-05$ & $2.70E-05$ & $0.2376$ & $0.2587$ & $0.2411$ & $0.2558$ & $1.66$\\
$1.00$ & $30$ & $-11$   & $1.15E-05$ & $1.99E-05$ & $0.1752$ & $0.1884$ & $0.4435$ & $0.4656$ & $1.60$\\
$1.00$ & $30$ & $-12$   & $9.72E-06$ & $2.17E-05$ & $0.1407$ & $0.1665$ & $0.5910$ & $0.5971$ & $1.66$\\
$1.00$ & $30$ & $-12.3$ & $3.21E-05$ & $5.53E-05$ & $0.1607$ & $0.2064$ & $0.5588$ & $0.5431$ & $2.12$\\
\hline
$1.00$ & $50$ &  $-6$   & $2.15E-06$ & $0.00E+00$ & $0.3220$ & {\nodata}& $0.0203$ & {\nodata}& $1.03$\\
$1.00$ & $50$ &  $-7$   & $8.30E-06$ & $5.05E-06$ & $0.3239$ & $0.3470$ & $0.0207$ & $0.0210$ & $1.39$\\
$1.00$ & $50$ &  $-8$   & $2.27E-05$ & $2.27E-05$ & $0.3625$ & $0.3891$ & $0.0227$ & $0.0272$ & $1.58$\\
$1.00$ & $50$ &  $-9$   & $1.62E-05$ & $1.94E-05$ & $0.2718$ & $0.2934$ & $0.1836$ & $0.1957$ & $1.53$\\
$1.00$ & $50$ & $-10$   & $1.09E-05$ & $1.42E-05$ & $0.2622$ & $0.2809$ & $0.2486$ & $0.2636$ & $1.46$\\
$1.00$ & $50$ & $-11$   & $7.94E-06$ & $1.19E-05$ & $0.2220$ & $0.2450$ & $0.3710$ & $0.3790$ & $1.43$\\
$1.00$ & $50$ & $-12$   & $2.07E-05$ & $3.27E-05$ & $0.2328$ & $0.2756$ & $0.4323$ & $0.4210$ & $1.78$\\
$1.00$ & $50$ & $-12.3$ & $2.53E-05$ & $4.69E-05$ & $0.2274$ & $0.2748$ & $0.4947$ & $0.4764$ & $1.96$\\
\hline
$1.25$ & $10$ &  $-6$   & $4.14E-07$ & $0.00E+00$ & $0.3184$ & {\nodata}& $0.0204$ & {\nodata}& $1.27$\\
$1.25$ & $10$ &  $-7$   & $1.92E-06$ & $1.62E-06$ & $0.3194$ & $0.3744$ & $0.0217$ & $0.0263$ & $1.62$\\
$1.25$ & $10$ &  $-8$   & $3.67E-06$ & $3.91E-06$ & $0.2983$ & $0.3637$ & $0.0834$ & $0.0876$ & $1.90$\\
$1.25$ & $10$ &  $-9$   & $9.27E-06$ & $1.06E-05$ & $0.2562$ & $0.3336$ & $0.1417$ & $0.1411$ & $2.33$\\
$1.25$ & $10$ & $-10$   & $1.91E-05$ & $2.18E-05$ & $0.2473$ & $0.3221$ & $0.1413$ & $0.1562$ & $2.70$\\
$1.25$ & $10$ & $-11$   & $2.97E-05$ & $3.61E-05$ & $0.2372$ & $0.3155$ & $0.1933$ & $0.2092$ & $3.01$\\
$1.25$ & $10$ & $-12$   & $3.22E-05$ & $5.33E-05$ & $0.1889$ & $0.2485$ & $0.3696$ & $0.4371$ & $3.77$\\
$1.25$ & $10$ & $-12.3$ & \multicolumn{7}{c}{\nodata}\\
\hline
$1.25$ & $30$ &  $-6$   & $3.82E-07$ & $0.00E+00$ & $0.3299$ & {\nodata}& $0.0203$ & {\nodata}& $1.23$\\
$1.25$ & $30$ &  $-7$   & $1.96E-06$ & $1.86E-06$ & $0.3837$ & $0.4380$ & $0.0224$ & $0.0273$ & $1.67$\\
$1.25$ & $30$ &  $-8$   & $3.84E-06$ & $4.16E-06$ & $0.3027$ & $0.3671$ & $0.0996$ & $0.1037$ & $1.92$\\
$1.25$ & $30$ &  $-9$   & $5.22E-06$ & $5.86E-06$ & $0.2730$ & $0.3400$ & $0.1305$ & $0.1359$ & $2.05$\\
$1.25$ & $30$ & $-10$   & $4.35E-06$ & $5.55E-06$ & $0.2423$ & $0.3060$ & $0.2380$ & $0.2480$ & $2.05$\\
$1.25$ & $30$ & $-11$   & $2.34E-06$ & $4.26E-06$ & $0.1676$ & $0.2202$ & $0.4733$ & $0.4931$ & $1.99$\\
$1.25$ & $30$ & $-12$   & $2.10E-06$ & $4.25E-06$ & $0.1421$ & $0.2124$ & $0.5747$ & $0.5725$ & $2.01$\\
$1.25$ & $30$ & $-12.3$ & $5.79E-06$ & $9.91E-06$ & $0.1858$ & $0.2708$ & $0.5548$ & $0.5401$ & $2.84$\\
\hline
$1.25$ & $50$ &  $-6$   & $4.16E-07$ & $0.00E+00$ & $0.3148$ & {\nodata}& $0.0205$ & {\nodata}& $1.26$\\
$1.25$ & $50$ &  $-7$   & $1.96E-06$ & $1.78E-06$ & $0.3374$ & $0.4221$ & $0.0218$ & $0.0269$ & $1.67$\\
$1.25$ & $50$ &  $-8$   & $3.69E-06$ & $4.01E-06$ & $0.3026$ & $0.3661$ & $0.1019$ & $0.1069$ & $1.89$\\
$1.25$ & $50$ &  $-9$   & $3.18E-06$ & $3.58E-06$ & $0.2803$ & $0.3415$ & $0.1498$ & $0.1583$ & $1.86$\\
$1.25$ & $50$ & $-10$   & $2.14E-06$ & $2.78E-06$ & $0.2584$ & $0.3133$ & $0.2523$ & $0.2663$ & $1.78$\\
$1.25$ & $50$ & $-11$   & $1.62E-06$ & $2.51E-06$ & $0.2210$ & $0.2820$ & $0.3930$ & $0.3970$ & $1.76$\\
$1.25$ & $50$ & $-12$   & $4.43E-06$ & $5.95E-06$ & $0.3387$ & $0.4160$ & $0.3356$ & $0.3264$ & $2.12$\\
$1.25$ & $50$ & $-12.3$ & $5.57E-06$ & $1.00E-05$ & $0.4400$ & $0.5262$ & $0.3346$ & $0.3169$ & $2.46$\\
\hline
$1.40$ & $10$ &  $-6$   & $1.81E-08$ & $0.00E+00$ & $0.3372$ & {\nodata}& $0.0213$ & {\nodata}& $1.64$\\
$1.40$ & $10$ &  $-7$   & $7.71E-08$ & $5.31E-08$ & $0.3087$ & $0.4603$ & $0.0218$ & $0.0272$ & $2.09$\\
$1.40$ & $10$ &  $-8$   & $1.64E-07$ & $1.83E-07$ & $0.2977$ & $0.5004$ & $0.1277$ & $0.1277$ & $2.52$\\
$1.40$ & $10$ &  $-9$   & $4.12E-07$ & $4.74E-07$ & $0.2622$ & $0.4732$ & $0.1536$ & $0.1521$ & $3.07$\\
$1.40$ & $10$ & $-10$   & $5.90E-07$ & $6.90E-07$ & $0.2543$ & $0.4068$ & $0.1591$ & $0.2250$ & $3.62$\\
$1.40$ & $10$ & $-11$   & $2.59E-06$ & $3.03E-06$ & $0.2332$ & $0.4598$ & $0.2077$ & $0.1967$ & $4.70$\\
$1.40$ & $10$ & $-12$   & \multicolumn{7}{c}{\nodata}\\
$1.40$ & $10$ & $-12.3$ & \multicolumn{7}{c}{\nodata}\\
\hline
$1.40$ & $30$ &  $-6$   & $1.78E-08$ & $0.00E+00$ & $0.3466$ & {\nodata}& $0.0208$ & {\nodata}& $1.62$\\
$1.40$ & $30$ &  $-7$   & $7.94E-08$ & $5.54E-08$ & $0.3073$ & $0.4574$ & $0.0217$ & $0.0271$ & $2.10$\\
$1.40$ & $30$ &  $-8$   & $2.02E-07$ & $2.02E-07$ & $0.3600$ & $0.5410$ & $0.0242$ & $0.0339$ & $2.51$\\
$1.40$ & $30$ &  $-9$   & $2.64E-07$ & $3.02E-07$ & $0.2817$ & $0.4891$ & $0.1488$ & $0.1497$ & $2.81$\\
$1.40$ & $30$ & $-10$   & $2.11E-07$ & $2.68E-07$ & $0.2509$ & $0.4643$ & $0.2338$ & $0.2325$ & $2.76$\\
$1.40$ & $30$ & $-11$   & $1.28E-07$ & $2.09E-07$ & $0.1968$ & $0.3986$ & $0.4235$ & $0.4325$ & $2.65$\\
$1.40$ & $30$ & $-12$   & $9.30E-08$ & $1.94E-07$ & $0.1189$ & $0.3213$ & $0.6100$ & $0.6179$ & $2.66$\\
$1.40$ & $30$ & $-12.3$ & \multicolumn{7}{c}{\nodata}\\
\hline
$1.40$ & $50$ &  $-6$   & $1.80E-08$ & $0.00E+00$ & $0.3354$ & {\nodata}& $0.0214$ & {\nodata}& $1.62$\\
$1.40$ & $50$ &  $-7$   & $8.09E-08$ & $5.69E-08$ & $0.3089$ & $0.4772$ & $0.0218$ & $0.0277$ & $2.07$\\
$1.40$ & $50$ &  $-8$   & $2.02E-07$ & $2.02E-07$ & $0.3600$ & $0.5385$ & $0.0241$ & $0.0336$ & $2.48$\\
$1.40$ & $50$ &  $-9$   & $1.90E-07$ & $2.18E-07$ & $0.2927$ & $0.4952$ & $0.1535$ & $0.1541$ & $2.60$\\
$1.40$ & $50$ & $-10$   & $1.21E-07$ & $1.64E-07$ & $0.2608$ & $0.4678$ & $0.2848$ & $0.2838$ & $2.48$\\
$1.40$ & $50$ & $-11$   & $6.83E-08$ & $1.25E-07$ & $0.2220$ & $0.4230$ & $0.4750$ & $0.4730$ & $2.40$\\
$1.40$ & $50$ & $-12$   & $2.44E-07$ & $3.15E-07$ & $0.3772$ & $0.6059$ & $0.3348$ & $0.3202$ & $4.20$\\
$1.40$ & $50$ & $-12.3$ & \multicolumn{7}{c}{\nodata}\\

\enddata

\tablenotetext{*}{$log\dot{M}$ of $-12.3$ stands for $\dot{M}=5\times 10^{-13}$}

\end{deluxetable}

\begin{deluxetable}{ccc|ccccccc}

\tablecaption{Complete Grid Results - Characteristics of The Outburst\label{tbl:outburst}}
\tabletypesize{\scriptsize}
\tablewidth{0pt}
\tablehead{
\multicolumn{3}{c}{Param. Comb.} &
\multicolumn{7}{c}{Outburst Characteristics} \\
\colhead{$M_{WD}$} & \colhead{$T_{WD}$} & \colhead{$log\dot{M}$} & 
\colhead{$v_{max}$} & \colhead{$v_{avg}$} & \colhead{$L_{4,max}$} & \colhead{$A$} &
\colhead{$t_{3,bol}$} & \colhead{$t_{m-l}$} &
\colhead{$P_{rec}$} \\
\colhead{} & \colhead{} & \colhead{} &
\colhead{$(km~s^{-1})$} &
\colhead{$(km~s^{-1})$} & \colhead{$(10^4~L_\odot)$} & \colhead{} &
\colhead{$(days)$} & \colhead{$(days)$} & 
\colhead{$(yr)$} 
}
\startdata
$0.40$ & $10$ &  $-9$   & $234$  & $149$  & $7.03$  & $15.6$ & $1.10E+04$ & $4.96E+01$ & $4.12E+05$\\
$0.40$ & $10$ & $-10$   & $203$  & $97$   & $11.90$ & $18.2$ & $8.06E+03$ & $6.19E+01$ & $5.62E+06$\\
$0.40$ & $10$ & $-11$   & $819$  & $462$  & $23.88$ & $21.0$ & $5.50E+04$ & $3.87E+01$ & $5.87E+07$\\
\hline
$0.65$ & $10$ &  $-6$   & $0$    & $0$    & $1.18$  & $4.2$  & $2.65E+04$ & $0.00E+00$ & $8.35E+00$ \\
$0.65$ & $10$ &  $-7$   & $0$    & $0$    & $1.48$  & $6.9$  & $5.67E+04$ & $0.00E+00$ & $2.45E+02$ \\
$0.65$ & $10$ &  $-8$   & $156$  & $122$  & $1.52$  & $9.4$  & $2.93E+04$ & $1.17E+03$ & $1.01E+04$ \\
$0.65$ & $10$ &  $-9$   & $2590$ & $2150$ & $4.76$  & $13.2$ & $3.83E+04$ & $2.64E+02$ & $1.61E+05$ \\
$0.65$ & $10$ & $-10$   & $4210$ & $2650$ & $13.70$ & $16.8$ & $3.23E+04$ & $1.17E+02$ & $2.55E+06$ \\
$0.65$ & $10$ & $-11$   & $1300$ & $623$  & $5.98$  & $18.0$ & $1.53E+04$ & $2.76E+01$ & $2.58E+07$ \\
$0.65$ & $10$ & $-12$   & $682$  & $216$  & $20.72$ & $20.7$ & $8.69E+03$ & $1.10E+02$ & $3.94E+08$ \\
$0.65$ & $10$ & $-12.3$ & $543$  & $230$  & $17.34$ & $20.9$ & $8.22E+02$ & $6.20E+02$ & $1.08E+09$ \\
\hline
$0.65$ & $30$ &  $-6$   & $0$    & $0$    & $1.17$  & $4.1$  & $2.82E+04$ & $0.00E+00$ & $8.63E+00$ \\
$0.65$ & $30$ &  $-7$   & $0$    & $0$    & $1.49$  & $6.9$  & $6.01E+04$ & $0.00E+00$ & $2.54E+02$ \\
$0.65$ & $30$ &  $-8$   & $139$  & $125$  & $1.58$  & $9.5$  & $3.22E+04$ & $1.22E+03$ & $1.02E+04$ \\
$0.65$ & $30$ &  $-9$   & $210$  & $156$  & $1.65$  & $12.0$ & $1.27E+04$ & $6.79E+02$ & $1.11E+05$ \\
$0.65$ & $30$ & $-10$   & $316$  & $195$  & $7.96$  & $16.2$ & $8.38E+03$ & $4.83E+02$ & $9.55E+05$ \\
$0.65$ & $30$ & $-11$   & $471$  & $189$  & $2.16$  & $15.2$ & $4.38E+02$ & $3.74E+01$ & $5.96E+06$ \\
$0.65$ & $30$ & $-12$   & $544$  & $268$  & $2.72$  & $15.4$ & $6.08E+02$ & $2.75E+01$ & $4.12E+07$ \\
$0.65$ & $30$ & $-12.3$ & $736$  & $297$  & $2.85$  & $15.5$ & $5.41E+02$ & $2.61E+01$ & $8.89E+07$ \\
\hline
$0.65$ & $50$ &  $-6$   & $0$    & $0$    & $1.14$  & $4.1$  & $2.98E+04$ & $0.00E+00$ & $8.94E+00$ \\
$0.65$ & $50$ &  $-7$   & $0$    & $0$    & $1.42$  & $6.9$  & $6.44E+04$ & $0.00E+00$ & $2.66E+02$ \\
$0.65$ & $50$ &  $-8$   & $159$  & $130$  & $1.60$  & $9.5$  & $4.38E+04$ & $1.17E+03$ & $1.06E+04$ \\
$0.65$ & $50$ &  $-9$   & $340$  & $240$  & $1.73$  & $12.0$ & $7.91E+03$ & $4.39E+02$ & $7.41E+04$ \\
$0.65$ & $50$ & $-10$   & $369$  & $208$  & $1.91$  & $13.6$ & $7.51E+03$ & $3.59E+02$ & $5.22E+05$ \\
$0.65$ & $50$ & $-11$   & $416$  & $179$  & $2.17$  & $13.7$ & $6.54E+03$ & $3.56E+02$ & $3.86E+06$ \\
$0.65$ & $50$ & $-12$   & $342$  & $176$  & $2.87$  & $16.0$ & $5.52E+03$ & $4.00E+01$ & $4.54E+07$ \\
$0.65$ & $50$ & $-12.3$ & $552$  & $487$  & $25.32$ & $16.5$ & $1.56E+03$ & $1.69E+01$ & $2.33E+08$ \\
\hline
$1.00$ & $10$ &  $-6$   & $0$    & $0$    & $2.96$  & $4.2$  & $2.43E+03$ & $0.00E+00$ & $2.05E+00$ \\
$1.00$ & $10$ &  $-7$   & $265$  & $240$  & $3.45$  & $6.9$  & $2.72E+03$ & $2.10E+02$ & $8.96E+01$ \\
$1.00$ & $10$ &  $-8$   & $351$  & $271$  & $3.26$  & $9.4$  & $1.35E+03$ & $1.27E+02$ & $2.06E+03$ \\
$1.00$ & $10$ &  $-9$   & $525$  & $256$  & $3.88$  & $12.0$ & $1.13E+03$ & $9.35E+01$ & $4.66E+04$ \\
$1.00$ & $10$ & $-10$   & $1920$ & $1180$ & $11.30$ & $15.7$ & $1.28E+02$ & $3.36E+01$ & $8.40E+05$ \\
$1.00$ & $10$ & $-11$   & $1250$ & $1063$ & $6.08$  & $17.0$ & $2.96E+03$ & $3.54E+01$ & $8.72E+06$ \\
$1.00$ & $10$ & $-12$   & $1230$ & $292$  & $9.80$  & $19.1$ & $4.19E+03$ & $5.53E+01$ & $9.28E+07$ \\
$1.00$ & $10$ & $-12.3$ & $1070$ & $295$  & $7.46$  & $18.7$ & $7.14E+03$ & $5.70E+01$ & $2.08E+08$ \\
\hline
$1.00$ & $30$ &  $-6$   & $0$    & $0$    & $2.96$  & $4.2$  & $2.53E+03$ & $0.00E+00$ & $2.10E+00$ \\
$1.00$ & $30$ &  $-7$   & $267$  & $237$  & $3.26$  & $6.9$  & $3.35E+03$ & $2.17E+02$ & $8.74E+01$ \\
$1.00$ & $30$ &  $-8$   & $355$  & $274$  & $3.28$  & $9.4$  & $1.32E+03$ & $1.57E+02$ & $2.03E+03$ \\
$1.00$ & $30$ &  $-9$   & $475$  & $324$  & $3.61$  & $12.0$ & $1.02E+03$ & $1.41E+02$ & $2.70E+04$ \\
$1.00$ & $30$ & $-10$   & $512$  & $344$  & $3.55$  & $14.4$ & $7.08E+02$ & $1.18E+02$ & $2.10E+05$ \\
$1.00$ & $30$ & $-11$   & $700$  & $446$  & $4.17$  & $15.3$ & $3.93E+02$ & $1.03E+01$ & $1.15E+06$ \\
$1.00$ & $30$ & $-12$   & $608$  & $414$  & $9.14$  & $16.5$ & $5.28E+02$ & $1.06E+01$ & $9.72E+06$ \\
$1.00$ & $30$ & $-12.3$ & $3520$ & $3490$ & $67.64$ & $18.5$ & $2.34E+02$ & $3.82E+00$ & $6.42E+07$ \\
\hline
$1.00$ & $50$ &  $-6$   & $0$    & $0$    & $2.97$  & $4.3$  & $2.60E+03$ & $0.00E+00$ & $2.15E+00$ \\
$1.00$ & $50$ &  $-7$   & $300$  & $239$  & $3.21$  & $6.8$  & $3.56E+03$ & $2.53E+02$ & $8.30E+01$ \\
$1.00$ & $50$ &  $-8$   & $360$  & $272$  & $3.60$  & $9.5$  & $2.51E+03$ & $2.40E+02$ & $2.27E+03$ \\
$1.00$ & $50$ &  $-9$   & $400$  & $328$  & $3.57$  & $12.0$ & $9.50E+02$ & $1.43E+02$ & $1.62E+04$ \\
$1.00$ & $50$ & $-10$   & $384$  & $302$  & $3.71$  & $14.0$ & $8.48E+02$ & $9.14E+01$ & $1.09E+05$ \\
$1.00$ & $50$ & $-11$   & $680$  & $471$  & $4.89$  & $14.3$ & $5.28E+02$ & $1.22E+02$ & $7.94E+05$ \\
$1.00$ & $50$ & $-12$   & $736$  & $597$  & $4.81$  & $14.7$ & $4.65E+02$ & $1.16E+01$ & $2.07E+07$ \\
$1.00$ & $50$ & $-12.3$ & $3170$ & $3083$ & $89.91$ & $18.4$ & $2.30E+02$ & $7.25E+00$ & $5.05E+07$ \\
\hline
$1.25$ & $10$ &  $-6$   & $0$    & $0$    & $4.57$  & $4.0$  & $3.11E+02$ & $0.00E+00$ & $4.14E-01$ \\
$1.25$ & $10$ &  $-7$   & $414$  & $346$  & $4.84$  & $6.5$  & $2.59E+02$ & $6.51E+01$ & $1.92E+01$ \\
$1.25$ & $10$ &  $-8$   & $1110$ & $568$  & $6.38$  & $9.3$  & $1.31E+02$ & $2.02E+01$ & $3.67E+02$ \\
$1.25$ & $10$ &  $-9$   & $1480$ & $413$  & $6.67$  & $11.9$ & $9.78E+01$ & $9.56E+00$ & $9.27E+03$ \\
$1.25$ & $10$ & $-10$   & $2230$ & $1940$ & $7.14$  & $14.5$ & $1.38E+02$ & $2.43E+01$ & $1.91E+05$ \\
$1.25$ & $10$ & $-11$   & $4240$ & $3857$ & $69.37$ & $19.1$ & $4.32E+02$ & $9.31E+00$ & $2.97E+06$ \\
$1.25$ & $10$ & $-12$   & $9680$ & $8267$ & $184.16$& $21.6$ & $4.44E+02$ & $5.86E+01$ & $3.22E+07$ \\
$1.25$ & $10$ & $-12.3$ & \multicolumn{7}{c}{\nodata}\\
\hline
$1.25$ & $30$ &  $-6$   & $0$    & $0$    & $4.52$  & $4.0$  & $2.46E+02$ & $0.00E+00$ & $3.82E-01$ \\
$1.25$ & $30$ &  $-7$   & $474$  & $373$  & $4.77$  & $6.5$  & $2.36E+02$ & $6.01E+01$ & $1.96E+01$ \\
$1.25$ & $30$ &  $-8$   & $734$  & $531$  & $4.78$  & $9.0$  & $1.62E+02$ & $3.22E+01$ & $3.84E+02$ \\
$1.25$ & $30$ &  $-9$   & $1030$ & $589$  & $5.14$  & $11.6$ & $1.25E+02$ & $3.14E+01$ & $5.22E+03$ \\
$1.25$ & $30$ & $-10$   & $1230$ & $678$  & $6.04$  & $14.3$ & $1.13E+02$ & $2.20E+01$ & $4.35E+04$ \\
$1.25$ & $30$ & $-11$   & $2110$ & $1264$ & $6.01$  & $15.4$ & $5.86E+01$ & $2.05E+00$ & $2.34E+05$ \\
$1.25$ & $30$ & $-12$   & $2710$ & $1696$ & $6.60$  & $15.8$ & $4.61E+01$ & $2.00E+00$ & $2.10E+06$ \\
$1.25$ & $30$ & $-12.3$ & $4260$ & $3668$ & $89.08$ & $18.8$ & $2.79E+00$ & $2.20E+00$ & $1.16E+07$ \\
\hline
$1.25$ & $50$ &  $-6$   & $0$    & $0$    & $4.53$  & $4.0$  & $3.15E+02$ & $0.00E+00$ & $4.16E-01$ \\
$1.25$ & $50$ &  $-7$   & $436$  & $355$  & $4.75$  & $6.5$  & $2.55E+02$ & $4.75E+01$ & $1.96E+01$ \\
$1.25$ & $50$ &  $-8$   & $807$  & $533$  & $4.61$  & $9.0$  & $1.58E+02$ & $4.30E+01$ & $3.69E+02$ \\
$1.25$ & $50$ &  $-9$   & $689$  & $485$  & $4.72$  & $11.5$ & $1.54E+02$ & $2.47E+01$ & $3.18E+03$ \\
$1.25$ & $50$ & $-10$   & $921$  & $542$  & $4.97$  & $14.1$ & $1.01E+02$ & $3.04E+01$ & $2.14E+04$ \\
$1.25$ & $50$ & $-11$   & $889$  & $658$  & $5.45$  & $14.3$ & $7.77E+01$ & $2.59E+01$ & $1.62E+05$ \\
$1.25$ & $50$ & $-12$   & $1350$ & $751$  & $6.30$  & $14.6$ & $6.91E+01$ & $2.70E+00$ & $4.43E+06$ \\
$1.25$ & $50$ & $-12.3$ & $3130$ & $1675$ & $12.25$ & $15.9$ & $4.45E+01$ & $2.08E+00$ & $1.11E+07$ \\
\hline
$1.40$ & $10$ &  $-6$   & $0$    & $0$    & $5.92$  & $3.4$  & $1.08E+01$ & $0.00E+00$ & $1.81E-02$ \\
$1.40$ & $10$ &  $-7$   & $1410$ & $681$  & $6.03$  & $5.9$  & $1.19E+01$ & $4.13E+00$ & $7.71E-01$ \\
$1.40$ & $10$ &  $-8$   & $3060$ & $1160$ & $9.27$  & $8.8$  & $5.02E+00$ & $1.57E+00$ & $1.64E+01$ \\
$1.40$ & $10$ &  $-9$   & $2850$ & $1020$ & $9.23$  & $11.3$ & $4.31E+00$ & $1.39E+00$ & $4.12E+02$ \\
$1.40$ & $10$ & $-10$   & $5270$ & $2020$ & $10.10$ & $13.9$ & $5.45E+00$ & $6.78E-01$ & $5.90E+03$ \\
$1.40$ & $10$ & $-11$   & $4660$ & $2817$ & $37.54$ & $17.5$ & $4.71E+01$ & $2.95E+00$ & $2.59E+05$ \\
$1.40$ & $10$ & $-12$   & \multicolumn{7}{c}{\nodata}\\
$1.40$ & $10$ & $-12.3$ & \multicolumn{7}{c}{\nodata}\\
\hline
$1.40$ & $30$ &  $-6$   & $0$    & $0$    & $5.86$  & $3.3$  & $1.11E+01$ & $0.00E+00$ & $1.78E-02$ \\
$1.40$ & $30$ &  $-7$   & $1160$ & $723$  & $5.99$  & $5.9$  & $1.19E+01$ & $3.64E+00$ & $7.94E-01$ \\
$1.40$ & $30$ &  $-8$   & $1760$ & $840$  & $6.14$  & $8.4$  & $9.35E+00$ & $4.63E+00$ & $2.02E+01$ \\
$1.40$ & $30$ &  $-9$   & $3760$ & $1800$ & $6.98$  & $11.0$ & $6.17E+00$ & $1.29E+00$ & $2.64E+02$ \\
$1.40$ & $30$ & $-10$   & $4490$ & $2120$ & $7.67$  & $13.6$ & $4.20E+00$ & $1.50E+00$ & $2.11E+03$ \\
$1.40$ & $30$ & $-11$   & $4710$ & $2738$ & $8.41$  & $15.3$ & $3.57E+00$ & $1.87E-01$ & $1.28E+04$ \\
$1.40$ & $30$ & $-12$   & $3310$ & $2312$ & $9.52$  & $16.1$ & $3.05E+00$ & $1.91E-01$ & $9.30E+04$ \\
$1.40$ & $30$ & $-12.3$ & \multicolumn{7}{c}{\nodata}\\
\hline
$1.40$ & $50$ &  $-6$   & $0$    & $0$    & $5.85$  & $3.3$  & $1.11E+01$ & $0.00E+00$ & $1.80E-02$ \\
$1.40$ & $50$ &  $-7$   & $1360$ & $812$  & $6.01$  & $5.8$  & $1.25E+01$ & $4.24E+00$ & $8.09E-01$ \\
$1.40$ & $50$ &  $-8$   & $1540$ & $833$  & $6.46$  & $8.4$  & $9.49E+00$ & $4.80E+00$ & $2.02E+01$ \\
$1.40$ & $50$ &  $-9$   & $3740$ & $1580$ & $7.18$  & $11.1$ & $6.03E+00$ & $1.76E+00$ & $1.90E+02$ \\
$1.40$ & $50$ & $-10$   & $4350$ & $1870$ & $7.67$  & $13.6$ & $4.71E+00$ & $1.17E+00$ & $1.21E+03$ \\
$1.40$ & $50$ & $-11$   & $3630$ & $1900$ & $7.74$  & $14.8$ & $3.45E+00$ & $1.67E+00$ & $6.83E+03$ \\
$1.40$ & $50$ & $-12$   & $3550$ & $2570$ & $9.31$  & $14.9$ & $3.81E+00$ & $3.13E-01$ & $2.44E+05$ \\
$1.40$ & $50$ & $-12.3$ & \multicolumn{7}{c}{\nodata}\\

\enddata

\end{deluxetable}

\begin{deluxetable}{lc|ccl|c|ccl}

\tablecaption{Grid Results - Maximum and Minimum Values of Nova Characteristics (for $0.65-1.40 \Msun$)\label{tbl:maxmin}}
\tablewidth{0pt}
\tablehead{
\colhead{} & \colhead{Max} & \multicolumn{3}{c}{Param. Comb.} & \colhead{Min} &
\multicolumn{3}{c}{Param. Comb.} \\
\colhead{Characteristic} & \colhead{Value} & \colhead{$M_{WD}$} & \colhead{$T_{WD}$} & \colhead{$log\dot{M}$} &  \colhead{Value} & \colhead{$M_{WD}$} & \colhead{$T_{WD}$} & \colhead{$log\dot{M}$}
}
\startdata

$m_{acc}$    & $5.40E-04$  &  $0.65$ & $10$ & $-12.3$  & $6.83E-08$ & $1.40$ & $50$ & $-11$\\
$m_{ej}$     & $6.66E-04$  &  $0.65$ & $10$ & $-12$    & $5.31E-08$ & $1.40$ & $10$ & $-7$\\
$Z_{ej}$     & $0.63$      &  $0.65$ & $30$ & $-12.3$  & $0.02$     & $1.00$ & $50$ & $-7$\\
$Y_{ej}$     & $0.61$      &  $1.40$ & $50$ & $-12$    & $0.12$     & $0.65$ & $30$ & $-12.3$\\
$X_{ej}$     & $0.65$      &  $0.65$ & $10$ & $-12.3$  & $0.06$     & $1.40$ & $30$ & $-12$\\
$T_{8,max}$  & $4.7$       &  $1.40$ & $10$ & $-11$    & $1.1$      & $0.65$ & $50$ & $-11$\\
$v_{max}$    & $5270$      &  $1.40$ & $10$ & $-10$    & $139$      & $0.65$ & $30$ & $-8$\\
$v_{avg}$    & $3860$      &  $1.25$ & $10$ & $-11$    & $122$      & $0.65$ & $10$ & $-8$\\
$L_{4,bol,max}$  & $90$    &  $1.00$ & $50$ & $-12.3$  & $1.5$      & $0.65$ & $10$ & $-8$\\
$M_{bol,max}$    & $-10.2$ &  $1.00$ & $50$ & $-12.3$  & $-5.7$     & $0.65$ & $10$ & $-8$\\
$A$          & $20.9$      &  $0.65$ & $10$ & $-12.3$  & $5.8$      & $1.40$ & $50$ & $-7$\\
$t_{3,bol}$  & $120~yr$    &  $0.65$ & $50$ & $-8$     & $3.05~d$   & $1.40$ & $30$ & $-12$\\
$t_{m-l}$    & $67~yr$     &  $0.65$ & $10$ & $-11$    & $0.68~d$   & $1.40$ & $10$ & $-10$\\
$P_{rec}$    & $1.08E+09~yr$ & $0.65$ & $10$ & $-12.3$ & $281~d$    & $1.40$ & $10$ & $-7$\\
\enddata

\end{deluxetable}

\begin{deluxetable}{lc|c|c}

\tablecaption{Observed vs. Calculated Nova Characteristics Ranges\label{obs-calc}}
\tablewidth{0pt}
\tablehead{
\colhead{Characteristic} & \colhead{Observed Ranges} &
\colhead{Observed Exceptions} & \colhead{Calculated Ranges}
}
\startdata

$M_{max}$ & $-6$ to $-9$     & $-10$ (V1500~Cyg)  & $-5.7$ to $-10.2$ \\
$A$       & $7 - 16$         & $19.3$ (V1500~Cyg) & $5.8 - 20.9$ \\
$t_3$     & $4 - 300$ days   & SymN:more          & $0.76~d - 67~yr$ \\
$m_{ej}$  & $1 - 30 (E-05 M_\odot)$ & RN:less   & $5.3E-08 - 6.6E-04$ \\
$Z_{ej}$  & $0.04 - 0.41$    & $0.86$ (V1370 Aql) & $0.02 - 0.63$ \\
$Y_{ej}$  & $0.21 - 0.48$    & $0.1$ (V1370 Aql)  & $0.12 - 0.60$ \\
$v_{exp}$\tablenotemark{*} & $350 - 2500~km~s^{-1}$ &
SymN:$\sim100~km~s^{-1}$ & $122 - 3860$ \\

\enddata
\tablenotetext{*}{average expansion velocities}

\end{deluxetable}

\begin{deluxetable}{c||cc|cc}

\tablecaption{Ejecta Composition - CO vs. ONe Models\label{tbl:CO-ONe}}
\tablewidth{0pt}
\tablehead{
\colhead{} & \multicolumn{2}{c}{$1.25,\ 10,\ -11$} &
\multicolumn{2}{c}{$1.40,\ 10,\ -9$}\\
\colhead{} & \colhead{$CO$} & \colhead{$ONe$} &  \colhead{$CO$} &
\colhead{$ONe$}
}

\startdata
$m_{ej}$   & $3.61E-05$ & $3.67E-05$ & $4.74E-07$ & $2.89E-07$\\
$y_{ej}$   & $0.3155$ & $0.3123$ & $0.4732$ & $0.5129$\\
$z_{ej}$   & $0.2092$ & $0.2179$ & $0.1521$ & $0.1508$\\
\hline
\tablenotemark{*}
$C^{12}$   & $3.353E-02$ & $1.970E-02$ & $2.430E-02$ & $1.143E-02$\\
$C^{13}$   & $3.136E-02$ & $2.089E-02$ & $2.485E-02$ & $8.148E-03$\\
$N^{14}$   & $6.545E-02$ & $3.086E-02$ & $9.392E-02$ & $4.706E-02$\\
$N^{15}$   & $6.660E-02$ & $3.321E-02$ & $4.191E-03$ & $1.147E-03$\\
$O^{16}$   & $1.165E-02$ & $9.241E-05$ & $2.092E-04$ & $1.402E-04$\\
$O^{17}$   & $6.157E-04$ & $7.729E-05$ & $1.342E-05$ & $8.500E-06$\\
$Ne$       & $3.384E-07$ & $6.579E-02$ & $1.372E-07$ & $1.358E-03$\\
$Na$       & $3.535E-09$ & $6.541E-04$ & $1.047E-10$ & $1.141E-06$\\
$Mg$       & $2.861E-08$ & $9.611E-03$ & $5.238E-09$ & $3.640E-05$\\
$Al^{26}$  & $2.098E-08$ & $7.929E-03$ & $6.355E-09$ & $4.258E-05$\\
$Al^{27}$  & $1.027E-08$ & $3.137E-03$ & $1.251E-09$ & $7.662E-06$\\
$\geq Si$  & $1.046E-07$ & $2.600E-02$ & $6.087E-07$ & $8.147E-02$\\
\enddata

\tablenotetext{*}{Heavy elements $Z_{ej}$ breakdown}

\end{deluxetable}

\end{document}